\documentclass[fleqn,usenatbib]{mnras}

\usepackage{float}
\usepackage{graphicx}	% Including figure files
\usepackage{amsmath}	% Advanced maths commands
\usepackage{amssymb}	% Extra maths symbols
\usepackage{multicol}        % Multi-column entries in tables
\usepackage{bm}		% Bold maths symbols, including upright Greek
\usepackage{pdflscape}	% Landscape pages
\usepackage{multirow,array}
\usepackage{subcaption}
\usepackage{xcolor}
\usepackage{booktabs}
\usepackage{pdflscape}
\usepackage{pifont}
\usepackage{array,xcolor,colortbl}

\renewcommand{\arraystretch}{1.1}
\newcolumntype{C}{>{\centering\columncolor{white}}p{0.022\textwidth}}

\newcounter{inst}
\title[Modelling Remnant AGNs]{Selecting and Modelling Remnant AGNs with Limited Spectral Coverage}

\author[B. Quici et al.]{Benjamin Quici$^{1}$\thanks{Email: benjamin.quici@icrar.org}, Ross J. Turner$^{2}$, Nicholas Seymour$^{1}$, Natasha Hurley-Walker$^{1}$,\and Stanislav S. Shabala$^{2,3}$, C.~H.~Ishwara-Chandra$^{4}$\\
$^{1}$International Centre for Radio Astronomy Research, Curtin University, Bentley, WA 6102, Australia\\
$^{2}$School of Natural Sciences, University of Tasmania, Private Bag 37, Hobart, 7001, Australia\\
$^{3}$ARC Centre of Excellence for All-Sky Astrophysics in 3 Dimensions (ASTRO 3D)\\
$^{4}$National Centre for Radio Astrophysics, TIFR, Post Bag No. 3, Ganeshkhind Post, 411007 Pune, India}

\date{Accepted XXX. Received YYY; in original form ZZZ.}

\pubyear{2022}

\usepackage{aas_macros}
\usepackage{hyperref} 
\hypersetup{colorlinks,citecolor=blue,linkcolor=blue,urlcolor=blue}

\def\app#1#2{%
  \mathrel{%
    \setbox0=\hbox{$#1\sim$}%
    \setbox2=\hbox{%
      \rlap{\hbox{$#1\propto$}}%
      \lower1.1\ht0\box0%
    }%
    \raise0.25\ht2\box2%
  }%
}
\def\approxprop{\mathpalette\app\relax}

\newcommand{\remnant}{J2253-34}
\newcommand{\miriad}{\textsc{miriad}}

\newcommand{\agns}{AGNs}

\begin{document}

%\begin{frontmatter}
\maketitle

\begin{abstract}
Quantifying the energetics and lifetimes of remnant radio-loud active galactic nuclei (\agns{}) is much more challenging than for active sources due to the added complexity of accurately determining the time since the central black hole switched off. Independent spectral modelling of remnant lobes enables the derivation of the remnant ratio, $R_\mathrm{rem}$, (i.e. `off-time/source age'), 
% thus reducing this complexity back to that of active sources
however, the requirement of high-frequency ($\gtrsim5$\,GHz) coverage makes the application of this technique over large-area radio surveys difficult. In this work we propose a new method, which relies on the observed brightness of backflow of Fanaroff-Riley type~II lobes, combined with the \emph{Radio AGN in Semi-Analytic Environments} (RAiSE) code, to measure the duration of the remnant phase. Sensitive radio observations of the remnant radio galaxy \remnant{} are obtained to provide a robust comparison of this technique with the canonical spectral analysis and modelling methods. We find that the remnant lifetimes modelled by each method are consistent; spectral modelling yields $R_\mathrm{rem} = 0.23\pm0.02$, compared to $R_\mathrm{rem} = 0.26\pm0.02$ from our new method. We examine the viability of applying our proposed technique to low-frequency radio surveys using mock radio source populations, and examine whether the technique is sensitive to any intrinsic properties of radio AGNs. Our results show that the technique can be used to robustly classify active and remnant populations, with the most confident predictions for the remnant ratio, and thus off-time, in the longest-lived radio sources ($>50$~Myr) and those at higher redshifts ($z > 0.1$).

\end{abstract}
% \rule{\linewidth}{.7pt}\vspace{5pt}

%\end{frontmatter}

\section{INTRODUCTION}
\label{sec:intro}

The supermassive black holes (SMBHs) residing at the heart of most galaxies \citep{1998AJ....115.2285M,2004ApJ...604L..89H,2009ApJ...698..198G}, play a profound role in the evolution of their host galaxies and large-scale (intergalactic) environments. Accretion onto the SMBH powers an active galactic nucleus (AGN), which, when radio-loud, drives a pair of jets comprising relativistic plasma. The jets inflate synchrotron-emitting lobes in the surrounding atmosphere, which act to heat up and expel the surrounding intergalactic gas, and eventually halt accretion onto the SMBH \citep[e.g. see reviews by;][]{2007ARA&A..45..117M,2012ARA&A..50..455F,2012NewAR..56...93A}. 
This \emph{jet mode} of AGN feedback is needed to quench star formation in the most massive galaxies \citep{2006MNRAS.365...11C}, and to suppress cooling flows in the cores of massive clusters \citep{2003MNRAS.344L..43F} at low redshifts \citep[$z\lesssim2-3$;][]{2012ARA&A..50..455F}. 

Prescriptions for these feedback mechanisms require more than just the energetics and lifetimes of the jetted outbursts \citep{2007MNRAS.377..142B}; the manner by which the jet power couples with the surrounding gas is equally as important \citep{2008MmSAI..79.1205T,2012NewAR..56...93A}. The lifetime of the active jet phase has a large influence on the mechanism coupling the jet power to the environment; e.g. through shock-heating driven by the global expansion of the lobes \citep{2012MNRAS.424.1346W}, versus the clearing of gas through the late buoyant rise of jet-inflated bubbles \citep{2001ApJ...554..261C}. This active lifetime also strongly maps to the radio source linear size \citep{1997MNRAS.286..215K}, which determines the cluster radius of interaction where the energy deposition takes place. The duration of a quiescent phase, in which there is either no or reduced jet activity, can additionally be constrained from the active lifetime when combined with measurements of the duty cycle. The duty cycle can be directly constrained for ``double-double radio galaxies'' \citep[DDRG;][]{2000MNRAS.315..371S} and some remnant radio galaxies \citep{2018MNRAS.476.2522T}, or estimated on a population level from the radio-loud fraction \citep[e.g.][]{2005MNRAS.362...25B,2019A&A...622A...1S}. Feedback mechanisms can therefore be examined by considering the spatial and temporal scales over which the energy deposition takes place; i.e. does the cluster environment have sufficient time to reach its equilibrium state before jet activity resumes \citep[e.g.][]{2015ApJ...806...59T}.  

The feedback mechanisms that regulate the supply of the surrounding gas are thought to give rise to an intermittency in jet activity, and the observed ``remnant'' and ``restarted'' radio galaxy populations \citep[e.g. see reviews by;][]{2015aska.confE.173K,2017NatAs...1..596M}; the latter representing those in which jet activity has renewed prior to the remnant lobes fading from view. Importantly, remnant radio galaxies offer valuable constraints on the active lifetimes of AGN jets, necessary towards our general understanding of AGN jet feedback. \citet{2020MNRAS.496.1706S} demonstrate that models describing the global distribution of jet lifetimes and duty-cycles can be inferred from the observed fractions of remnant and restarted radio sources (relative to active ones). A direct measurement of the active jet lifetime can be made by modelling the dynamical evolution of remnant lobes \citep{2018MNRAS.476.2522T}; the same is not true of the age derived for active radio lobes, which corresponds only to the age of the source at the time of observation, rather than the total lifetime of the jet. For these reasons, methods to select and model remnant radio galaxy lobes need to be robust.

Methodologies for selecting remnant radio galaxies are explored by many authors, particularly since the advent of the Low Frequency Array \citep[LOFAR;][]{2013A&A...556A...2V}. The overall consensus, as first proposed by \citet{2016A&A...585A..29B}, is that a combination of morphological- and spectral-based methods are needed to select a representative sample of remnant radio galaxies. 
%  I think that the sentence should be re-stated; the emission region(s) need to fade at certain frequency in order for the spectrum to curve/steepen. I think that a more precise observation is that such selection may miss the very young remnants, (which have breaks at very high frequencies), usually not covered by the data et at hand and/or undetectable due to low surface brightness at those frequencies.

It is possible to select remnants based on an ultra-steep \citep[e.g. $\alpha\geq1.2$;][]{1987MNRAS.227..695C}, or highly curved \citep[e.g. $\Delta\alpha\geq0.5$;][]{2011A&A...526A.148M} radio spectrum criterion. These methods rely on sufficient fading (ageing) at the highest observing frequency, and will therefore miss recently switched-off remnants in which the break frequency lies beyond this frequency; this is confirmed based on observed and mock radio source populations \citep[e.g see;][]{2017MNRAS.471..891G,2017A&A...606A..98B,2018MNRAS.475.4557M,2020MNRAS.496.1706S}. 
% Selecting remnants based only on an ultra-steep \citep[e.g. $\alpha\geq1.2$;][]{1987MNRAS.227..695C} or highly curved \citep[$\Delta\alpha\geq0.5$;][]{2011A&A...526A.148M} radio spectrum criterion, results in a biased sample of aged remnants due to the lobes often fading faster than the spectra steepen \citep{2017MNRAS.471..891G}; this is confirmed based on observed and mock radio source populations \citep[e.g see;][]{2017A&A...606A..98B,2018MNRAS.475.4557M,2020MNRAS.496.1706S}. 
The bias associated with these spectral-based methods can be mitigated through the inclusion of a morphological-based selection, where remnant lobes are selected based on diffuse, low surface-brightness, amorphous lobes \citep[e.g. Blob 1;][]{2016A&A...585A..29B}, however such methods may be less effective for recently switched-off remnants \citep[e.g. 3C~28;][]{2015MNRAS.454.3403H}. 

Upon cessation of the jets, a bright radio core is expected to disappear rapidly ($<0.1\,$Myr). The selection of remnant radio galaxies can therefore also be made based on the absence of compact radio emission from the core \citep[e.g;][]{2018MNRAS.475.4557M,2021PASA...38....8Q}, however we note the recent results of \citet{2021A&A...653A.110J} who demonstrate the co-existence of faint radio emission from the core together with ultra-steep remnant radio lobes for a handful of objects. Their results suggest radio emission from the core, alone, may not necessarily determine whether the radio galaxy is active, and that high-frequency ($\nu\gtrsim5$\,GHz) observations are needed to confirm whether radio galaxy lobes are truly remnant.

High frequency radio observations are also critical to modelling the energetics of both active and remnant \agns{} \citep{2018MNRAS.474.3361T}. Ageing of the lobe plasma results in a spectral steepening due to the preferential radiating of high-energy electrons \citep{1962SvA.....6..317K,1970ranp.book.....P}, and is characterised by an optically-thin break frequency that can be related (given a lobe magnetic field strength) to the plasma spectral age. However, a major challenge in modelling the dynamical evolution of remnant \agns{} comes from the unknown time since the jets switched off. \citet{2018MNRAS.476.2522T} show that this problem can be resolved by modelling the fractional duration of the remnant phase, based on the observed steepening in the integrated lobe spectrum, reducing the required model complexity to that of active radio sources. However, this method requires a well-sampled radio spectrum, where high-frequency ($\nu\gtrsim5$\,GHz) observations are often needed to properly characterise the observed spectral curvature. 

% The overall picture painted here is that high-frequency ($\nu\gtrsim5$\,GHz) observations are essential for the selection and modelling of remnant lobes. 

Sensitive, wide-area radio surveys conducted with telescopes such as LOFAR; the Karoo Array Telescope \citep[MeerKAT;][]{2016mks..confE...1J}; the Australian Square Kilometre Array Pathfinder \citep[ASKAP;][]{2007PASA...24..174J}; and eventually the Square Kilometre Array (SKA), are expected to offer an unprecedented view of the radio galaxy life-cycle. In principle, such surveys enable the compilation of large samples of active, remnant and restarted radio sources, which will offer invaluable constraints on mechanisms describing the jet-triggering and AGN feedback. However, it is unrealistic to expect matching, high-frequency ($\nu\gtrsim5$\,GHz) sky coverage which, at present, is needed to robustly confirm the status of remnant lobes \citep{2021A&A...653A.110J}, and tightly constrain their energetics \citep{2018MNRAS.476.2522T}. 

% In the advent of these new-generation telescopes, producing images with high spatial resolution and excellent surface brightness sensitivity is becoming standard. It is with this consideration that we explore a new method that uses the observed surface brightness distribution to measure the off-time of remnant radio radio galaxies. 

In this work, we therefore explore whether measurements of the off-times of remnant lobes can be made based on their surface brightness distributions. To do this, we collect new data, presented in Sect.~\ref{sec:data}, for the remnant radio galaxy \remnant{} \citep[see;][]{2021PASA...38....8Q}, which are needed to conduct various spectral and dynamical modelling of the source. In Sect.~\ref{sec:spectral_modelling}, we apply an established spectral ageing method to provide an independent measurement of the off-time in \remnant{}. In Sect.~\ref{sec:dynamical_modelling}, we develop a new dynamical model-based method that uses the observed lobe backflow to measure the off-times in remnant radio galaxies. Here, we apply our method to \remnant{}, and validate the estimate of its off-time against that derived previously from its radio spectrum. In Sect.~\ref{sec:mock_populations}, we use mock radio source populations, simulated at 150\,MHz and 1.4\,GHz, to examine whether the off-times in a wide range of remnants can be constrained by this method. Final discussions and conclusions are summarised in Sect.~\ref{sec:conclusion}.

% we present new radio data collected for the remnant radio galaxy \remnant{}. In Sect.~\ref{sec:spectral_modelling}, we use these data to model the spectral ageing of the lobes, which are needed to independently verify our proposed method. In Sect.~\ref{sec:dynamical_modelling}, we develop attributes to parameterise the backflow of FR-II lobes based on the observed surface brightness distribution of the lobes, and use this to model the energetics of \remnant{}; here, we verify our results against the established method of \citet{2018MNRAS.476.2522T}. In Sect.~\ref{sec:mock_populations}, we extend our method to mock radio source populations simulated at 150\,MHz and 1.4\,GHz; here, our results inform the confidence with which the energetics can be constrained, and subsequently the ability to classify remnant lobes. Final discussions and conclusions are summarised in Sect.~\ref{sec:conclusion}.

The $\Lambda$CDM concordance cosmology with $\Omega_M$ = 0.3, $\Omega_\Lambda=0.7$ and $H_0$ = 70\,km\,s$^{-1}$\,Mpc$^{-1}$ \citep{2011ApJS..192...18K} is assumed throughout the paper.

\section{DATA}
\label{sec:data}
\begin{figure}
    \centering
    \includegraphics[width=\linewidth]{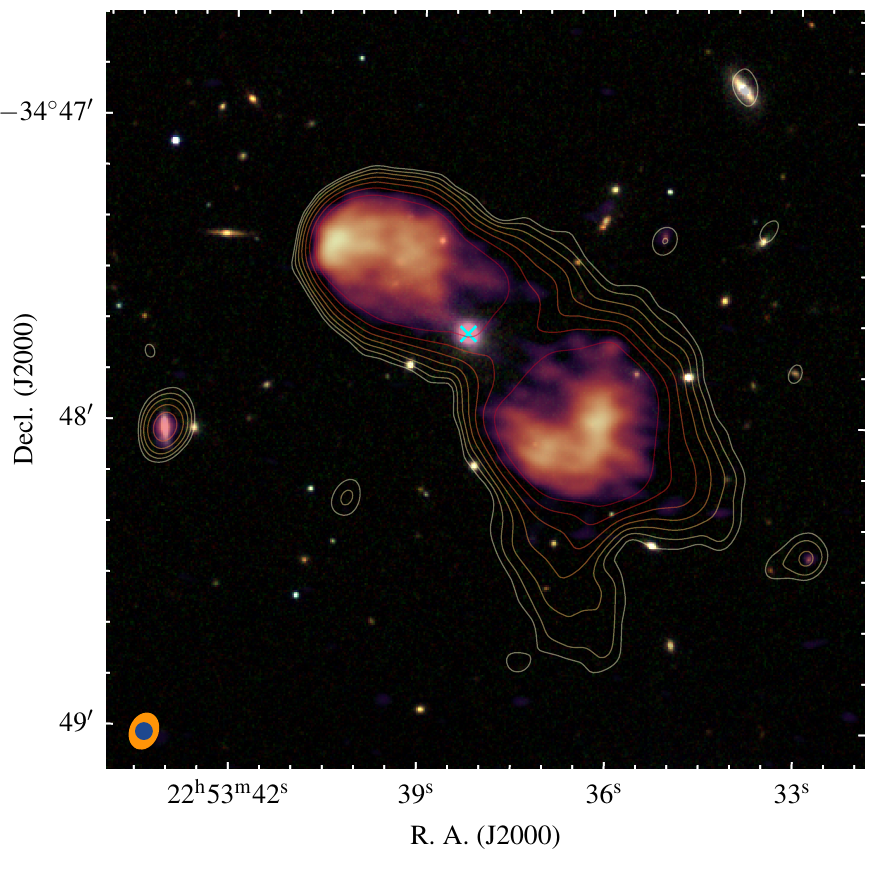}
    \caption{The remnant radio galaxy \remnant{} and its intergalactic environment. The background image is an RGB composite constructed from mid-infrared $K_s$ (red), $H$ (green) and $J$ (blue) bands, respectively \citep[e.g. see;][]{2021PASA...38....8Q}. The host galaxy is indicated by a cyan-coloured `x' marker. A 5.5\,GHz view of \remnant{} with $4''$ resolution is superimposed onto the background. Coloured contours show the 1.42\,GHz radio emission (`sub-band~3' in Sect.~\ref{sec:data:meerkat}), with levels increasing uniformly in log-space by $0.2$\,dex between [$3,80$]\,$\times\sigma$. Ellipses in the lower-left corner denote the shape of the restoring beam at 1.42\,GHz (orange) and 5.5\,GHz (blue).}
    \label{fig:remnant_image}
\end{figure}

In this section, we describe the observations taken as part of this work, or larger observing projects, for \remnant{}, and the methods used to image these data. We make use of new data collected by the Australia~Telescope~Compact~Array \citep[ATCA;][]{1992JEEEA..12..103F}, MeerKAT, and the upgraded Giant~Meterwave~Radio~Telescope (uGMRT), together with several ancillary data products. A multi-wavelength view of \remnant{} using these new data can be seen in Fig.~\ref{fig:remnant_image}. All measurements of the integrated flux density are reported in Tab.~\ref{tab:integrated_flux}. 

\subsection{ATCA data}
\begin{table*}
\centering
\caption{Summary of the new ATCA observations in different configurations described in Sect.~\ref{sec:atca}. In order, columns denote: the principal investigator; the project code; the ATCA configuration; the date of observation, each conducted during the year 2021; the total integration time; the range of shortest to longest baseline lengths; and the largest recoverable scales at 5.5\,GHz and 9\,GHz. We note, B$_{\mathrm{max}}$ quoted for all but the 6C configuration excludes baselines formed with the $\sim$6\,km outrigger antenna (CA06), due to the gap in $(u,v)$ coverage.}
\label{tab:observing}
\begin{tabular}{lccccccc}
\toprule
\multirow{2}{*}{PI} & \multirow{2}{*}{Project code} & \multirow{2}{*}{Config.} & Date & $t$ & B$_{\mathrm{min}}$ -- B$_{\mathrm{max}}$ & $\theta_{\mathrm{max, 5.5}}$ & $\theta_{\mathrm{max, 9}}$ \\
& &  & dd/mm & (h) & (m) & ($'$) & ($'$)\\
\midrule
\multirow{5}{*}{Benjamin Quici} & C3335 & H75  & $03/06$ & 8 & 31 -- 89 & 6.0 &3.7 \\
& C3335 & H168 & $22/07$ & 5.5 & 61 -- 185 & 3.1 & 1.9\\
& C3402 & 750C & $14/02$ & 12  & 46 -- 704 & 4.1 & 2.5\\
& C3402 & 1.5A & $03/01$ & 12  & 153 -- 1469 & 2.4 & 1.5\\
& C3402 & 6C   & $(12,15)/11$ & 10  & 337 -- 5939 & 1.2 & 0.7\\
\bottomrule
\end{tabular}
\end{table*}
    \label{sec:atca}
    Pointed observations of \remnant{} were awarded through a series of observing proposals (Tab.~\ref{tab:observing}).
    % , with the motivation of constraining the energetics of the plasma through high-frequency observations (both high- and low- spatial resolution). 
    Observations were conducted within the $4$\,cm--band, which allows simultaneous observing over two 2\,GHz-wide bands which were centered at 5.5\,GHz and 9\,GHz. The collection and reduction of these data are presented below.
    
    To produce images with high spatial-resolution and sensitivity to large-scale structure, coverage of the $(u,v)$ plane was maximised by observing in various ATCA configurations (see Tab.~\ref{tab:observing}). Baselines ranged between 31--5939\,m, ensuring an appropriate sampling of the largest angular scale of \remnant{} ($\theta\sim100''$). The combined integration time on the target was 39.5\,hours. 
    % Observations all followed the same observing strategy. 
    At the beginning of each observation, we observed PKS~B1934-638 for 10\,minutes to derive bandpass, gain and flux calibration solutions. 12\,minute scans of \remnant{} were interleaved with two-minute scans of PKS~B2255-282; a source used to calibrate the time-varying gains.
    
    Data reduction was performed using the Multichannel Image Reconstruction, Image Analysis and Display \citep[\miriad{};][]{1995ASPC...77..433S}, and closely followed the method outlined by \cite{2015MNRAS.454..952H,2020MNRAS.491.3395H}; we used the primary scan to solve for the shape of the bandpass, the secondary scan to calibrate for the time-varying gains and the polarization leakages, and then transferred the solutions to \remnant{} where the gain solutions were averaged over a two-minute interval (the length of each phase calibrator scan). Throughout all these steps, channels known to be contaminated by radio frequency interference (RFI) were automatically flagged.
    Calibrated data-sets associated with each unique observation, with equal weighting, into a single data-set. To improve the imaging dynamic range, three rounds of phase-only self calibration were conducted using a \texttt{Robust~0.0} Briggs weighting \citep{1995AAS...18711202B}, corresponding to an image resolution of approximately $3''\times6''$. For each round, gains were solved over 2,~1,~and~0.5\,minute time intervals, respectively. We refer to this data below as the `calibrated data-set', and discuss the synthesis of our science-ready images in Sects.~\ref{sec:atca_lowres_cx}~\&~\ref{sec:atca_highres_cx}. For these data, we assume a $3\%$ calibration uncertainty in the ATCA flux density scale \citep{Reynolds1994}.
    
    % To create a final data-set ready for imaging, calibrated data-sets collected in the H75, H168, 750C and 1.5A configurations were concatenated together using the \texttt{uvcat} task. In this way, the raw sensitivity and coverage of the $(u,v)$\,plane were both maximised. Three rounds of phase-only self calibration were conducted in order to improve the dynamic range in the image. For each round, gains were solved over 2\,minute, 1\,minute and 0.5\,minute time intervals, respectively. We refer to this data below as the `calibrated data-set', and discuss the synthesis of our science-ready images in Sections~\ref{sec:atca_highres_cx}~\&~\ref{sec:atca_lowres_cx}.

    \subsubsection{Low-resolution imaging}
    \label{sec:atca_lowres_cx}
    We first produced images with lower spatial resolution to improve constraints on the high-frequency, integrated, spectral energy distribution~(SED). This was done by splitting each 2-GHz band of the calibrated data-set into four 512\,MHz sub-bands. Each sub-band was imaged in \textsc{miriad} using a \texttt{Robust~0.0} Briggs weighting, with an additional $30''$ Gaussian taper on the visibilities so as to reduce contribution and noise from the longer baselines. The data were CLEANed to an initial depth of 3$\sigma$, after which a deeper CLEAN to 1$\sigma$ was performed around pixels already in the model. Images were corrected for attenuation of the primary beam. 
    
    To extract the photometry, we used \textsc{polygon\_flux}\footnote{\url{https://github.com/nhurleywalker/polygon-flux}}, which uses the shape of the restoring beam to evaluate a background-subtracted integrated flux density corresponding to a user-defined region (e.g. see \citealt{2019PASA...36...48H}). Measurement uncertainties were calculated by multiplying the RMS noise by the square-root number of beams corresponding to the region volume, which we added in quadrature with the absolute calibration uncertainty.
    
    \subsubsection{High-resolution imaging}
    \label{sec:atca_highres_cx}
    We further produced images with high spatial resolution so we could map the radio spectrum across the lobes (in conjunction with our data at other frequencies). To maximise signal-to-noise, we imaged the entire 2\,GHz bandwidth available in each wide-band. To produce these images, we used a \texttt{uniform} Briggs weighting and applied a $8''$ taper to the visibilities; although it was possible to improve the spatial resolution beyond $\theta_{\mathrm{FWHM}}=8''$, this corresponded to the best possible circular resolution that could be achieved across all radio images presented in this work. To demonstrate this in Fig.~\ref{fig:remnant_image}, we present a higher-resolution view of \remnant{} at 5.5\,GHz by instead applying a $4''$ taper to the visibilities. The data were CLEANed to an initial depth of 3$\sigma$, after which a deeper CLEAN to 1$\sigma$ was performed around pixels already in the model. After correcting images for the frequency-dependent primary beam attenuation, they achieved a mean root~mean~square (RMS) sensitivity of 11\,$\mu$Jy\,beam$^{-1}$ and 13\,$\mu$Jy\,beam$^{-1}$ at 5.5\,GHz and 9\,GHz, respectively. We refer to these images as `high-resolution ATCA images'.

\subsection{MeerKAT data}
\label{sec:data:meerkat}
%As part of a 2020 call for observing proposals, a proposal requesting a total of
We obtained 3.5\,hours of time to observe \remnant{} on the MeerKAT radio telescope (PI: B. Quici, project code: MKT~20212). Observations took place in the `L-band' (0.89--1.71\,GHz), thus offering sensitive, high spatial-resolution observations at an intermediate frequency range. To collect the data, we used PKS~1934-638 to calibrate the bandpass, and PKS~2259-37 to calibrate the time-varying gains. The MeerKAT bandpass response is stable to within 3\% over 2--3\,hours. As such, the observations were conducted in two `blocks' by visiting the bandpass calibrator twice; once at the beginning of the observation, and again 1.5\,hours later. Each bandpass calibrator scan lasted 10\,minutes. Observations of \remnant{} were carried out by interleaving a two\,minute gain calibrator scan with a 15.5\,minute target scan.
% To observe \remnant{}, we alternated between a two\,minute gain calibrator scan and a 15.5\,minute target scan, ensuring to end each block on a two\,minute gain calibrator scan. 
In total, the target received 2.6\,hours of exposure time.
Using the calibrated measurement set made available from the South African Radio Astronomy Observatory (SARAO)~Web~Archive\footnote{The SARAO Web Archive is found at \url{https://archive.sarao.ac.za/}}, we performed a single round of RFI flagging using AOFlagger \citep{2012A&A...539A..95O}. We identified four frequency ranges as having RFI-clean, unflagged visibilities, and we refer to these as `sub--bands' (see Tab.~\ref{tab:integrated_flux}).
% By inspecting the calibrated amplitudes as a function of channel, we identified a significant fraction of unflagged RFI missed by the standard calibration pipeline. As such we performed an independent round of RFI flagging using AOFlagger \citep{2012A&A...539A..95O}. The following four frequency ranges, which we label as sub--bands, were identified as having RFI-clean, unflagged visibilities: $0.89-0.932$\,GHz (sub--band 1); $0.96-1.08$\,GHz (sub--band 2); $1.345-1.498$\,GHz (sub--band 3); $1.638-1.668$\,GHz (sub--band 4). In this way, approximately $42\%$ of the total bandwidth was used.

Imaging was conducted using WSClean~\citep{2014ascl.soft08023O}. Each sub--band was imaged independently by blindly CLEANing the data down to a depth of 3$\sigma$, followed by a deeper CLEAN down to a $1\sigma$ level around the model. Imaging of sub--bands 1--4 was performed using an image weighting of \texttt{robust~-1.0}, \texttt{-0.6}, \texttt{0.0} and \texttt{+0.25}, respectively.
% We imaged sub--band~1 using a \texttt{robust~-1.0} weighting; sub--band~2 using a \texttt{robust~-0.6} weighting; sub--band~3 using a \texttt{robust~0.0} weighting; and sub--band~4 using a \texttt{robust~+0.25} weighting. 
In this way, all but the lowest sub--band achieved an image resolution of $\theta_{\mathrm{FWHM}}\lesssim8''$. 
% Due to the relatively large frequency range of sub--bands~2~\&~3, the bandwidth was split into two equal parts which allowed the CLEAN algorithm to search for peaks in the multi-frequency (MF) image, but perform the subtraction from individual parts. Both parts were joined at the end to produce a sensitive MF image. 
All imaging was performed using the \texttt{multiscale} option, which enables multi-scale deconvolution \citep{2017MNRAS.471..301O}. Each sub--band was then corrected for the frequency-dependent primary beam attenuation. Finally, images associated with sub--bands~2--4 were convolved to a circular 8$''$ resolution, in order to match the resolution in the high-resolution ATCA images. The image associated with sub--band~1 was excluded from this final step, as the restoring beam exceeded $8''$. A 5\% calibration uncertainty is assumed throughout this work.% We used the NRAO VLA Sky Survey \citep[NVSS; ][]{1998AJ....115.1693C}, centered at 1.4\,GHz, to estimate the uncertainty in the MeerKAT flux density scale. This was done by comparing the integrated flux density of \remnant{} measured in sub--band~3; the band closest in frequency to NVSS. By measuring integrated flux density ratio between the two data-sets, we measured $S_\mathrm{NVSS}/S_\mathrm{sub-band 3}=0.98$ and thus concluded our MeerKAT data was consistent with NVSS. As such, we assigned a 3\% uncertainty in the MeerKAT flux density scale, which is the value quoted for NVSS. 
Following the method as described in Sect.~\ref{sec:atca_lowres_cx}, the integrated flux density of \remnant{} (and associated measurement uncertainty) was measured from each sub--band using the same user-defined polygon. 
%  These values, as well their corresponding image properties, are reported in Table~\ref{tab:integrated_flux}. 
\subsection{uGMRT data}
\label{sec:gmrt}
We finally made use of two low-frequency, high-resolution radio images, created using data collected with the upgraded Giant~Meterwave~Radio~Telescope (uGMRT), which contained \remnant{} within their fields of view. Observations were conducted in band-3 ($250 - 500$\,MHz; proposal code 37\_057, PI: Ross Turner), and in band-4 ($550-900$\,MHz; proposal~code 35\_022, PI:~C.~H.~Ishwara~Chandra). The data was reduced using the CAPTURE Pipeline \citep{2021ExA....51...95K}, which uses the Common Astronomy Software Applications \citep[\textsc{casa};][]{2007ASPC..376..127M} package. 
After the initial gain calibration, the data was channel averaged to keep the bandwidth smearing negligible.
Using a Briggs weighting of robust~\texttt{0.0}, the channel averaged source data was subjected to four rounds of phase-only self-calibration followed by 4 more rounds of amplitude and phase self-calibration. A final band-3 image, centered at $417$\,MHz, was produced using a robust~\texttt{0.0} weighting, resulting in a native resolution of $\sim8''\times5''$. A final band-4 image, centered at $682$\,MHz, was produced using a robust~\texttt{0.0} weighting, resulting in a native resolution of $\sim5''\times3''$. The resolution in each image was convolved to a circular beam of $\theta=8''$, achieving a local RMS of $52\mu$Jy\,beam$^{-1}$ and $43\mu$Jy\,beam$^{-1}$ at 417\,MHz and 682\,MHz, respectively.

\subsection{Ancillary data}
To fill out the sampled integrated radio spectrum of \remnant{}, particularly at frequencies not covered by our new data, we take several measurements previously compiled by \citet{2021PASA...38....8Q}. We take their integrated flux densities reported at: 119~and~154\,MHz, measured from the GaLactic and Extragalactic All-sky Murchison Widefield Array \citep[GLEAM;][]{2017MNRAS.464.1146H} survey; 887\,MHz, measured from the Evolutionary Map of the Universe \citep[EMU;][]{2011JApA...32..599N}; 1.4\,GHz, measured from the NRAO VLA Sky Survey \citep[NVSS;][]{1998AJ....115.1693C}, and; 2~and~2.868\,GHz, measured using pointed observations collected in the $LS$-band ($\lambda=4$\,cm) of ATCA (see Tab.~\ref{tab:integrated_flux}).

% \begin{figure}
%     \centering
%     \includegraphics[width=\linewidth]{J23-34.pdf}
%     \caption{The remnant radio galaxy \remnant{} and its intergalactic environment. The background image is an RGB composite constructed from mid-infrared $K_s$ (red), $H$ (green) and $J$ (blue) bands, respectively \citep[e.g. see;][]{2021PASA...38....8Q}. The host galaxy is indicated by a cyan-coloured `x' marker. A 5.5\,GHz view of \remnant{} is superimposed onto the background; this image is made following the method presented in Sect.~\ref{sec:atca_highres_cx}, however to show the source with maximal spatial resolution, we instead apply a $4''$ circular taper to the visibilities. Coloured contours show the 1.42\,GHz radio emission (`sub-band~3' in Sect.~\ref{sec:data:meerkat}), with levels increasing uniformly in log-space by $0.2$\,dex between [$3,80$]\,$\times\sigma$. Ellipses in the lower-left corner denote the shape of the restoring beam at 1.42\,GHz (orange) and 5.5\,GHz (blue).}
%     \label{fig:remnant_image}
% \end{figure}

\section{SPECTRAL MODELLING OF REMNANT LOBES}
\label{sec:spectral_modelling}
Before we develop a new method to measure the off-time in remnant radio sources, we must ensure that our result can be verified against an established method. The off-time can be derived from the rapid steepening of their high-frequency spectrum due to the cessation of the jet activity and the subsequent lack of freshly shock-accelerated synchrotron-emitting electrons. In Sect.~\ref{sec:synchrotron_emissivity}--\ref{sec:field_independence}, we outline the theory of several literature-established spectral ageing models, and discuss their applications to radio lobe spectra. In Sect.~\ref{sec:remnant_modelling}, these models are applied to a combination of low- and high resolution radio observations, in order to tightly parameterise the spectral ageing observed for \remnant{}. Further, in Sect.~\ref{sec:synchrofit}, we briefly describe a \textsc{python} package we developed as part of this work to model synchrotron spectra for a diverse range of applications in astrophysics.

\subsection{Synchrotron emissivity}
\label{sec:synchrotron_emissivity}
The total emissivity of a synchrotron-emitting plasma,~$J(\nu)$, can be found by integrating the single-electron emissivity \citep[e.g. $j(\nu)$;][]{2011hea..book.....L} over the distribution of electron energies, $N(E)$; and the probabilistic distributions of the magnetic field strength, $p_B$, and pitch angle, $p_\xi$ \citep[Equation~4~of;][]{2013MNRAS.433.3364H}
\begin{equation}
    \label{eqn:synchrotron_emissivity}
    \resizebox{.92\hsize}{!}{$J(\nu) =  \int_{0}^{\infty} \int_{0}^{\pi}  \int_{0}^{\infty} \frac{\sqrt{3}Be^2\sin\xi}{8\pi^2\epsilon_0cm_e}F(x)N(E)p_\xi p_B dE d\xi dB$, }
\end{equation}
where $B$ gives the magnetic field strength; $\xi$ the pitch angle; $c$ the speed of light; $e$ the electron charge; $\epsilon_0$ the permittivity of free space; and $m_e$ the electron mass. \citet{1979rpa..book.....R} define $F(x)$ as the shape of the single-electron radiation spectrum, where $x$ represents a dimensionless parameterisation of frequency, magnetic field strength, and electron energy, $E$, through:
\begin{equation}
    \label{eqn:x}
    x = \frac{\nu}{\nu_\mathrm{b}} = \frac{4\pi m_\mathrm{e}^3c^4\nu}{3eE^2B\sin\xi},
\end{equation}
where $\nu_\mathrm{b}$ is the break~frequency above which (i.e. at higher frequencies) the spectrum steepens due to energy losses associated with the plasma ageing. Importantly, for a given magnetic field strength, the break frequency is related to the plasma spectral age through:
\begin{equation}
    \label{eqn:spectau}
    \tau  = \frac{\upsilon B^{1/2}}{B^2 + B_\mathrm{IC}^2}\big(\nu_\mathrm{b}(1+z)\big)^{-1/2},
\end{equation}
where $B_\mathrm{IC} = 0.318(1+z)^2$\,nT gives the magnetic field equivalent of the cosmic microwave background at redshift $z$, and $\upsilon$ is a constant of proportionality (see Equation~4 of \citealt{2018MNRAS.474.3361T}).

% t_{\mathrm{active}} + t_{\mathrm{remnant}}

Parameterisation of meaningful attributes, such as $\nu_\mathrm{b}$, demands appropriate treatment of $N(E)$, $p_B$, and $p_\xi$. %Several models are presented in the literature for the ageing of the electron energy distribution, $N(E)$. Of particular interest here, impulsive injection models, such as the Jaffe-Perola \citep[JP;][]{1973A&A....26..423J}, describe the synchrotron spectrum arising from a lobe section with equal-age electrons, whilst the continuous injection model \citep[CI;][]{1962SvA.....6..317K,1970ranp.book.....P} describes a broader region comprising mixed-age plasma. We described these two spectral ageing models shortly. %Meanwhile, it is typical to assume an isotropic distribution for the pitch angle, which sets $p_\xi=\frac{1}{2}\sin\xi$, and a locally-uniform magnetic field, such that $p_B=1$. 
% Regardless of the injection duration, these models both must make an assumption for the pitch angle and magnetic field strength probabilistic distributions.
\citet{1962SvA.....6..317K} initially proposed a model with no pitch angle scattering, giving $p_\xi = 1$, however a more realistic model assuming an isotropic distribution for the pitch angle was subsequently suggested by \citet{1973A&A....26..423J}, giving $p_\xi=\frac{1}{2}\sin\xi$; we assume the latter probabilistic distribution for the pitch angle in this work. Meanwhile, it is common to assume a locally-uniform magnetic field, such that $p_B=1$, however, this assumption is likely not physical due to the turbulent mixing of lobe plasma. \citet{1991MNRAS.253..147T} provide an alternative treatment for a turbulent (non-uniform) magnetic field, in which the field strength is drawn from a Maxwell-Boltzmann distribution; here, $p_B$ is satisfied by
\begin{equation}
    p_B = \sqrt{\frac{2}{\pi}} \frac{B^2\exp(-B^2/2a^2)}{a^3},
\end{equation}
where $a=B_0/\sqrt{3}$ for mean magnetic field strength, $B_0$ \citep{2013MNRAS.433.3364H}. %The spectral ageing models and assumption for the local magnetic field distribution can be paired to create a range of sub-models; 
%These are referred to as either the ``standard''~(uniform $B$) or ``Tribble''~(Maxwell-Boltzmann distribution $B$) forms of each spectral ageing model. %(e.g. ``standard-CI'' and ``Tribble-CI'', shortened to ``CI'' and ``TCI'' respectively). %These are explored in Sect.~\ref{sec:specmodels}. 
These assumptions can be paired together with an appropriate treatment of the electron energy distribution of the underlying plasma, $N(E)$ (described below), to create a range of spectral ageing models; 
these are referred to as either the ``standard''~(uniform $B$) or ``Tribble''~(Maxwell-Boltzmann distribution $B$) forms of each spectral ageing model.

%Finally, in the case of a uniform magnetic field, we can recast Equation~\ref{eqn:synchrotron_emissivity} to a double integral following the method of \citet{2018MNRAS.474.3361T} for the JP model, and \citet{2018MNRAS.474.3361T,2018MNRAS.476.2522T} for the CI models; we present this simplification for both the JP (Sect.~\ref{sec:JP}) and CI (Sect.~\ref{sec:CI}) models as below. \citet{2018MNRAS.474.3361T} discuss differences between the standard and Tribble forms of each spectral ageing model (their Sect.~2.3), which we expand upon in Sect.~\ref{sec:field_independence}.

%\subsection{Synchrotron spectral ageing models}
\label{sec:specmodels}
% Due to the preferential radiating of high-energy electrons \citep{1962SvA.....6..317K,1970ranp.book.....P}, the ageing of radio lobes introduces a steepening into their radio spectra. An optically thin break-frequency, $\nu_{\mathrm{b}}$, marks the frequency above which the steepening due to the plasma ageing can be observed, and encodes the spectral age of the underlying plasma population. If the observed radio source is a remnant, the spectrum will steepen additionally due to the absent supply of fresh (high-energy) plasma; \citet{2018MNRAS.476.2522T} parameterise these losses by the ``remnant ratio’’, $R_{\mathrm{rem}}$, which encodes the fractional duration of the remnant phase (``active~age/source~age’').
\subsubsection{Impulsive injection models}
\label{sec:JP}
The radio spectrum arising from an impulsively-injected electron population is modelled as an ensemble of equal-age electrons [$t,t+dt$] injected at $t=0$ \citep{1962SvA.....6..317K,1973A&A....26..423J}. Electrons are injected with an initial power-law distribution of energies, e.g $N(E) = N_0E^{-s}$, where the power-law energy index, $s$, is related to the injection spectral index, $\alpha_{\mathrm{inj}}$, through $s = 2\alpha_{\mathrm{inj}} + 1$. Synchrotron and inverse-Compton (IC) radiative losses, the latter due to up-scattering of Cosmic Microwave (CMB) photons, occur as the electron packet ages; the resulting effect on the energy distribution is well understood \citep[e.g. see;][]{1970ranp.book.....P,2011hea..book.....L}. In the case of a uniform magnetic field, we follow the method of \citet{2018MNRAS.474.3361T} to remove the dependence on $B$ by recasting $N(E)$ to:
\begin{equation}
    \label{eqn:JP_NE}
  N(x) = x^{-1/2}
%   x**(-1./2)*(np.sqrt(x) - 10**(x_crit/2))**(injection_index - 2)
  \begin{cases}
    (x^{1/2} - \iota^{1/2} )^{s-2}  & x > \iota \\
    0 & x < \iota
  \end{cases}
\end{equation}
where $\iota(\nu,\xi) = \nu/(\nu_{\mathrm{b}}\sin\xi)$ assuming an isotropic distribution for the pitch angle, as proposed by \citet{1973A&A....26..423J}. Here, the total synchrotron emissivity is thus evaluated as: 
\begin{equation}
    \label{eqn:Jnu_JP_reduced}
    \mathcal{J}(\nu) =  J_0\,\nu^{(1-s)/2}\int_{0}^{\pi/2}\sin^{(s+3)/2}\xi\int_0^\infty F(x) N(x) ~ dx\,d\xi,
    % \int_{0}^{\infty} \int_{0}^{\pi}  \int_{0}^{\infty} j(\nu)\,N(E)\,p_\xi\,p_B \quad dE\,d\xi\,dB,
\end{equation}
where $J_0$ acts as a frequency-independent constant. In this work, we refer to this impulsively-injected spectral ageing model as the ``standard'' Jaffe-Parola (or JP) model. 
Importantly, parameterising the shape of this standard JP spectrum in this way only requires knowledge of the injection index and break frequency. By comparison, the shape of a ``Tribble'' Jaffe-Parola (or TJP) model spectrum additionally requires knowledge of $B$.

Considering some small region across a radio galaxy lobe (e.g. bounded by the FWHM of a restoring beam), it is reasonable to suspect that the plasma contained here will have been injected at approximately the same age. This means that the associated radio spectrum is likely that of the JP model, however we note the results of \citet{2018MNRAS.473.4179T}, \citet{2020MNRAS.491.5015M} and \citet{2021MNRAS.508.5239Y,2022MNRAS.511.5225Y} who show that mixing of plasma ages may violate this assumption.

\subsubsection{Continuous Injection models}
\label{sec:CI}
The above assumption of an impulsively-injected packet of electrons breaks down when considering the integrated spectrum of a radio AGN, which comprises multiple, or on-going, such injections throughout the lifetime of the source. A continuous injection model describes a radio spectrum arising from a time-averaged population of continuously-injected electron packets; that is, a mixed-age plasma population with ages ranging within [$0, \tau$]. In the simpler and more common case, the CI model assumes continuous injection via an active jet throughout the entire lifetime of the source (e.g, the `CI-on' or CIJP\footnote{Here, the `JP' reflects only the isotropic distribution for the pitch angle proposed by \citet{1973A&A....26..423J}, not their assumption of an impulsively-injected plasma.} model). To model remnant AGNs, \citet{1994A&A...285...27K} present an extension to the CI model, known as the `CI-off' or KGJP model, by instead assuming some fraction of the total source age, $R_\mathrm{rem} = t_\mathrm{rem}/\tau$, is spent in a remnant phase; we refer to $R_\mathrm{rem}$ as the remnant ratio, where the total source age,~$\tau$, is given by the sum of the active and remnant durations, $\tau = t_{\mathrm{on}} + t_{\mathrm{rem}}$. As with the JP model, the effect of ageing on the plasma energy distribution is well understood \citep[e.g. see;][]{2011hea..book.....L}. In the case of a uniform magnetic field, the recast form of $N(E)$ is given by \citep[Equation~3 of;][]{2018MNRAS.476.2522T}
\begin{equation}
\mathcal{N}(x) = x^{-1/2}
\begin{cases}
\resizebox{!}{7.2pt}{$(x^{-1/2} - \zeta^{1/2})^{s-1} - (x^{1/2} - \iota^{1/2})^{s-1}$} & x > \iota  \\ 
\resizebox{!}{7.2pt}{$(x^{1/2} - \zeta^{1/2})^{s-1}$} & \zeta \leq x \leq \iota \\
0 & x < \zeta 
\end{cases},
\end{equation}
where $\zeta(\nu, \xi) = \iota(\nu, \xi)R_\mathrm{rem}^2$. The total synchrotron emissivity of a CI spectrum is thus derived by:
\begin{equation}
    \label{eqn:Jnu_CI_reduced}
    \mathcal{J}(\nu) =  J_0\,\nu^{-s/2}\int_{0}^{\pi/2}\sin^{(s+4)/2}\xi\int_0^\infty F(x) N(x) ~ dx\,d\xi,
\end{equation}
In the limiting case where $R_\mathrm{rem}=0$, the CI-off model reduces to the simpler CI-on form. For simplicity, we thus refer to the general form as the CI model. In this way, parameterising the shape of a standard CI spectrum requires only the injection index, break frequency, and remnant ratio. As with the TJP model, the shape of a ``Tribble'' Continuous Injection (or TCI) spectrum additionally requires knowledge of $B$. 

Due to their assumption of a mixed-age plasma, the CI models are relevant for deriving the plasma spectral age from the integrated spectra of radio lobes \citep[e.g;][]{2007A&A...470..875P,2011A&A...526A.148M,2012A&A...545A..91S,2016A&A...585A..29B,2018MNRAS.476.2522T,2019PASA...36...16D}. However, it is worth mentioning that the CI model assumes no temporal evolution in the magnetic field, which we know not to be true for expanding lobes. Using synthetic radio source populations, for which all radiative losses and temporal evolution of the magnetic field were considered, \citet{2018MNRAS.474.3361T} demonstrated that the radio emissivity reflects only the most recent $\sim10\%$ of the magnetic field history. Importantly, their results showed that the spectral age derived using the break frequency, parameterised by the CI model, together with the present-time magnetic field strength, was in agreement with the dynamical age. As such, we can be confident that the quantities parameterised by the CI model are physically meaningful. 

\subsection{Spectral model fitting}
\label{sec:synchrofit}
We now discuss our approach taken to numerically implement the spectral ageing model described in the previous section, and to fit these spectra to observed data points. The code we developed to achieve this is expected to be highly valuable to the broader astrophysics community (e.g. synchrotron spectra from active galaxies, supernovae remnants, etc.), and so we have released it as the publicly-available \textbf{synchro}tron~spectral~\textbf{fit}ter~(\textsc{synchrofit})\footnote{\url{https://github.com/synchrofit}} \textsc{python} library.
Principally, \textsc{synchrofit} offers a suite of routines that: 1) implement the JP and CI spectral ageing models described in Sect.~\ref{sec:JP}~and~\ref{sec:CI}, and 2) fit these models to real radio data. 

For computational efficiency, the standard spectral ageing models are evaluated by numerically integrating over the electron energy and pitch angle, following the reduced dimensionality parameterisation of $J(\nu)$, first presented by \citet[][JP]{2006ApJ...648..148N}, \citet[][CI-on]{2018MNRAS.474.3361T} and \citet[][CI-off]{2018MNRAS.476.2522T}. The Tribble spectral ageing models are evaluated in a similar fashion, however involve an additional numerical integration over the magnetic field strength, and as such incur a considerable cost to the computation time. An injection index and break frequency are required to parameterise the shape of the JP spectrum, with an additional normalisation to scale the fit appropriately. The shape of a CI spectrum is additionally parameterised by the remnant ratio, noting that $R_\mathrm{rem}=0$ in the case of the CI-on model. Tribble forms of the spectral ageing models are parameterised similarly, however additionally require the redshift and magnetic field strength. 

Fitting of these models is conducted using a grid-search informed by an adaptive maximum likelihood algorithm. Discrete samples in $s$, $\log{\nu_b}$ and $R_\mathrm{rem}$ are created by uniformly sampling their allowed ranges, which are used to construct a three-dimensional parameter grid. A spectral model is evaluated for each unique parameter combination, where a corresponding normalisation is optimised following $S_0 = \sum^n_{i=1} (S_{\mathrm{o_i}}\times\sigma_i) - \sum^n_{i=1} (S_{\mathrm{p_i}}\times\sigma_i) /\sum^n_{i=1} \sigma_i $, where $n$ is the number of spectral measurements, $S_\mathrm{o_i}$ and $S_\mathrm{p_i}$ are the observed and predicted flux densities in logarithmic space, and $\sigma_i$ is the measurement uncertainty for the $i$th datum. The quality of fit is evaluated using the lowest value of the Akaike information criterion \citep[AIC;][]{1974ITAC...19..716A} following $\text{AIC} = 2k - 2\ln\mathcal{L}$, where $k$ is the number of model parameters\footnote{For the standard JP and CI-on models, $k=3$. For the standard CI-off models, $k=4$. In the case of the Tribble forms, $k$ increases by 1 due to the magnetic field.}. The likelihood function, $\mathcal{L}$, is defined through:
\begin{equation}
\label{eqn:likelihood}
\mathcal{L}(\mu, \sigma) = \prod_{i=1}\frac{1}{\sqrt{2\pi}\sigma_i}\exp\Big(-\frac{(x_i-\mu_i)^2}{2\sigma_i^2}\Big)    
\end{equation}
 which, for each unique measurement, $i$, takes into consideration the observed value, $x_i$, measurement uncertainty, $\sigma_i$, and the corresponding predicted model value $\mu_i$. Across all grid realisations, the set of parameters providing the optimal spectral fit are defined as those that minimise the AIC. The relative quality of each spectral fit is found using $p_i = e^{(\mathrm{AIC}_i - \mathrm{AIC}_\mathrm{best})/2}$, where $p_i$ is the probability that the $i^{th}$ spectral model provides a better fit to the observations than the selected `best' model. Provided the marginal probability density functions (PDFs) in each model parameter are pseudo-Gaussian, the fitted uncertainty in each parameter is taken as the standard deviation, and the peak in the PDF as the approximate location of the global maximum. In order to increase the precision of the best estimate in each free parameter, a grid of points is then re-sampled around the global maximum; in this work, we repeat this over three iterations.
%  In order to increase the best estimate precision in each free parameter, a grid of points is then re-sampled about the peak probable value in each free parameter, in order to increase precision around the global maximum in the likelihood; this adaptive approach ensures the entire landscape of the likelihood is sampled within the bounds. 
We note, Gaussianity may not necessarily always hold for these PDFs, as this will depend upon the spectral fit, hence we make sure to inspect these throughout this work.
 
%  We note, Gaussianity may not necessarily always hold for these PDFs, as this will depend upon the spectral fit. Importantly, these PDFs where-ever applicable, and use these to ensure convergence occurs over a global maximum in the likelihood. To efficiently sample the entire parameter space, the grid search is performed over three iterations, each time further constraining the allowed parameter ranges about the most probable values in order to converge on the true spectral fit. 
 % provided the PDFs are pseudo-Gaussian, we approximate them using a 

\subsection{Effect of inhomogeneity in the local magnetic field}
\label{sec:field_independence}
    \begin{figure}
    \centering
    \includegraphics[width=\linewidth]{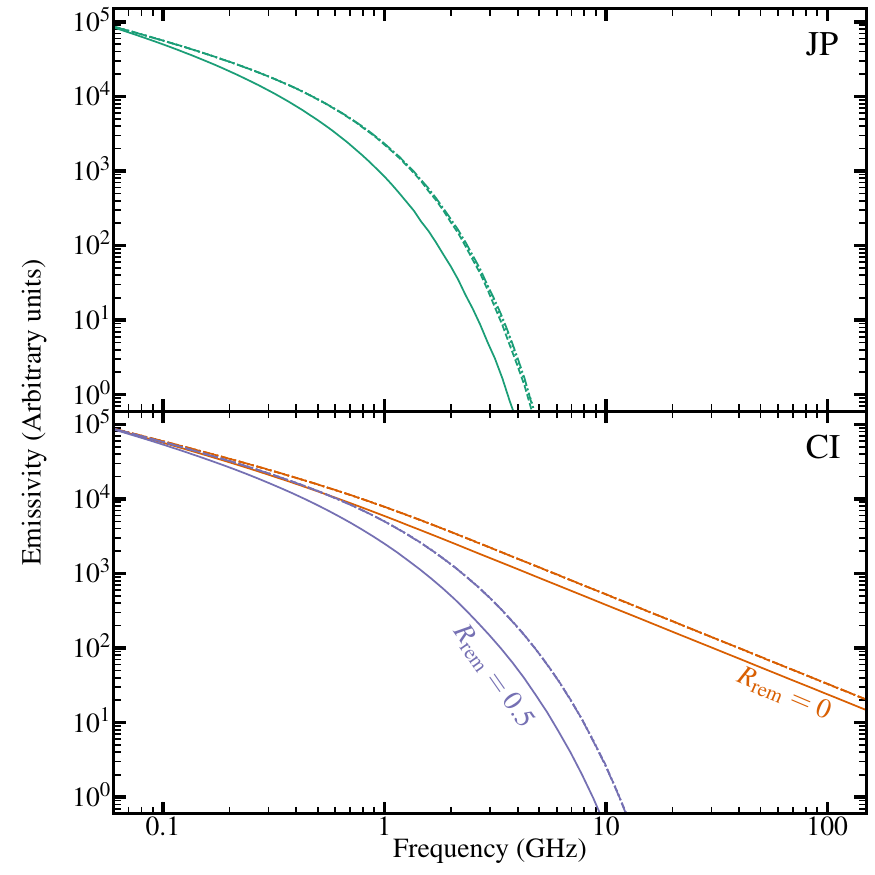}
    \caption{Comparison between standard (solid lines) and Tribble forms (broken lines) of the JP (top panel) and CI (bottom panel) spectra. \textsc{synchrofit} is used to simulate radio spectra for $s=2.2$ and $\nu_\mathrm{b}=2.1$\,GHz. Tribble models are shown for $B=0.1$\,nT (dot-dashed), $B=1$\,nT (dotted), and $B=10$\,nT (dashed). CI models are shown for $R_\mathrm{rem}=0$ (orange) and $R_\mathrm{rem}=0.5$ (purple). We note that the Tribble models almost exactly overlay each other and hence can not be distinguished.}
    \label{fig:standard_vs_tribble}
    \end{figure}
    
The standard form of the spectral ageing models have a considerable computational advantage over the (likely more physically accurate) Tribble forms in our implementation (due to the lower number of numerical integrations). We therefore investigate if the standard form can be used to reasonably approximate the Tribble form for either spectral ageing model, or for any range of magnetic field strengths.

Both the standard and Tribble forms of the spectral ageing models offer a physically-meaningful parameterisation of observed radio spectra. Due to the consideration of a variable magnetic field, the shape of the expected synchrotron spectrum differs to that in which the magnetic field is uniform. For example, \citet{2013MNRAS.433.3364H} show a significant departure of the TJP model from the standard form, notably due to the sharper spectral turnover around the break frequency in the TJP spectrum. \citet{2018MNRAS.474.3361T} explore the consequence of using the standard spectral ageing models to parameterise simulated Tribble spectra; their results demonstrate an over-estimation of the break frequency, which appears to be more appreciable for the JP model as compared to the CI model. 
% The implication here is that a source of uncertainty on the modelled spectral parameters can be introduced if .

The assumption of a locally uniform magnetic field will be at least somewhat violated in radio AGN lobes due to turbulent flow; the Maxwell-Boltzmann magnetic field distribution of the Tribble spectral ageing models may provide a more realistic description, however this in turn requires knowledge of $B_0$ which, unless independent X-ray data is available, would require dynamical modelling.

As such, we seek to explore the dependence of the shape of the Tribble spectral ageing models on $B_0$. We use \textsc{synchrofit} to simulate standard and Tribble forms of the JP and CI models, assuming $s=2.2$, and $\nu_\mathrm{b}=2.1$\,GHz. Tribble models are simulated for $B_0=0.1$\,nT, $1$\,nT and $10$\,nT, in order to cover a broad range of acceptable values. These are shown in Fig.~\ref{fig:standard_vs_tribble}.

%Importantly, several outcomes of this analysis are relevant to this work. 
As a consistency check with \citet{2013MNRAS.433.3364H}, \citet{2013MNRAS.435.3353H} and \citet{2018MNRAS.474.3361T}, we reassuringly find that integrating over a Maxwell-Boltzmann magnetic field modifies the shape of the predicted spectra to that in which the magnetic field is uniform. However, our results show that the shape of the predicted Tribble spectra are largely insensitive to values of the mean magnetic field strength. We find that electron packets with magnetic strengths in the exponential tail of the Maxwell-Boltzmann distribution (i.e.~$B > \sqrt{{3}/{2}}\, B_0$) contribute the vast majority of the total synchrotron emission at all radio frequencies. %That is, it is primarily the shape of the exponential tail in the magnetic field probability density function that encodes the shape of the synchrotron spectrum. 
Importantly, the mean magnetic field strength has no effect on the shape of this exponential portion of probability density function, albeit modifies its normalisation.
As a result, the shape of the radio spectrum remains largely unaffected for moderate changes in the mean magnetic field.% though the flux density scaling across the spectrum \citet[e.g.][]{2013MNRAS.433.3364H}.
The implication here is it is possible to accurately parameterise the injection index, break frequency, and remnant fraction (in the case of the CI models), using the Tribble spectral ageing models without explicitly requiring knowledge of $B_0$; an informed guess of the mean magnetic field strength is sufficient (e.g. $B_0\sim0.1$\,-\,$10$\,nT).

\subsection{Modelling the radio spectrum of \remnant{}}
\label{sec:remnant_modelling}
\begin{figure*}
\centering
\includegraphics[width=0.95\linewidth]{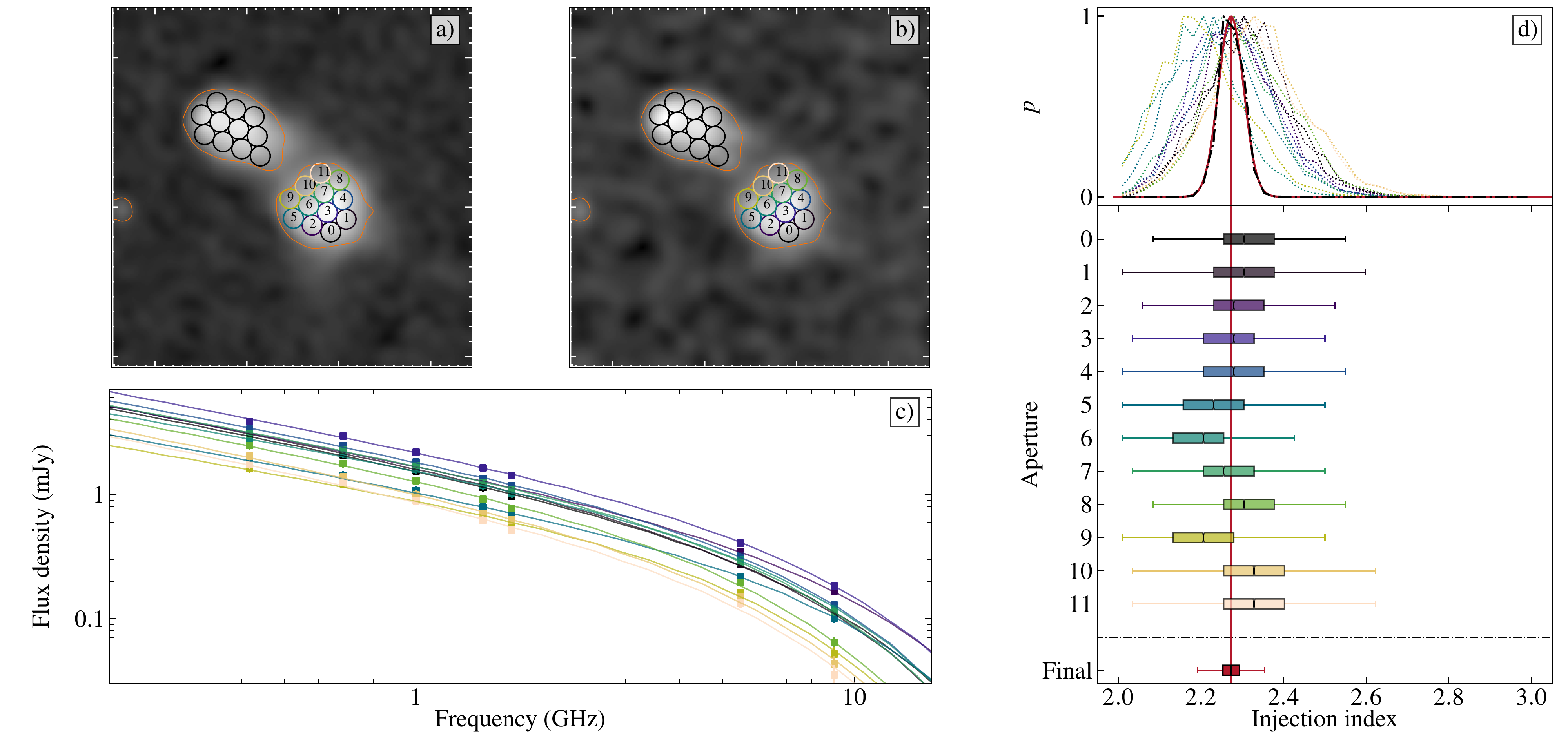}
% \caption{Results of jointly fitting for the injection index using multiple apertures for the south-western lobe of the remnant radio galaxy \remnant{}. Panels \emph{a)} and \emph{b)} show the observed radio emission at 1.32\,GHz and 5.5\,GHz; a solid dark-blue contour denotes the 5$\sigma$ brightness of the source at 9\,GHz. Each image is also overlaid with circular apertures of 8$''$ (the FWHM of the restoring beam) to show the regions from which the radio spectra are extracted. Each aperture spectrum is independently fit by the TJP model ($B_0=0.1$\,nT) where the injection index and break frequency are free parameters; the raw spectra, and their corresponding optimised spectral models, are presented in panel \emph{c)}. A precise measurement of the injection index is then found by considering that which offers the optimal global fit to all aperture-spectra; this is presented in panel \emph{d)}, and is summarised as follows. The probability density function (PDF) corresponding to each aperture spectral fit is marginalised for the injection index; these are represented by the dotted curves, and are also shown by the box and whisker plots, where the box is constrained by the $25^{\mathrm{th}}$\,-\,$75^{\mathrm{th}}$ percentiles. Each PDF is then multiplied together to optimise the injection index across the lobes; this is shown by the solid black curve, which is well approximated by a 1D~Gaussian shown in dark red. }
\caption{The injection index optimised for \remnant{}. Panels a) and b) show the observed radio emission at 1.32\,GHz and 5.5\,GHz, respectively; each image is presented on a square-root stretch, and is normalised between the minimum/maximum pixel values. A solid dark-orange contour denotes the 5$\sigma$ brightness of the source at 9\,GHz. Circular markers of an 8$''$ aperture (the FWHM of the restoring beam) are used to represent independent regions from which a radio spectrum is extracted and used as a constraint on the injection index. Apertures overlaid on the south-western lobe are colour-coded and numbered based on aperture number, and are highlighted in panels c) and d); to avoid clutter in the figure we do not plot the spectral fits for apertures in the north-eastern lobe. Panel c) demonstrates fits of the TJP model ($B_0=0.1$\,nT) optimised to their associated spectrum. Panel d) demonstrates the probability density function associated with each spectral fit, marginalised with respect to the injection index; these are also represented by the box and whisker plots, where the box is constrained by the $25^{\mathrm{th}}$\,-\,$75^{\mathrm{th}}$ percentiles. A global optimisation for the injection index is found by multiplying the PDFs associated with each lobe (dot-dashed black curve), which is then approximated by a 1D Gaussian (solid red curve) in order to characterise the peak probable value and uncertainty: $s~=~2.25\,\pm\,0.03$ (or alternatively, the injection spectral index; $\alpha_{\mathrm{inj}}=0.625\,\pm\,0.015$).}

% PDFs associated with each lobe are multiplied through together (solid black curve), and  is By multiplying each PDF associated with both lobes (solid black curve), which is approximated by a 1D Gaussian (thin red curve) to provide a final estimate of the injection index: $s~=~2.25\,\pm\,0.03$ (or alternatively, the injection spectral index; $\alpha_{\mathrm{inj}}=0.625\,\pm\,0.025$).
% }
\label{fig:injection}
\end{figure*}

We are now equipped with the tools needed to derive a robust measurement of the off-time from our observations \remnant{}. However, we choose not to directly fit the integrated spectrum with the CI-off model as the three parameters are somewhat correlated, and may be misfitted if our observations include moderate systematic uncertainties. Instead, we independently constrain the injection index by considering the spectra arising from narrow regions across the lobes; it is reasonable to assume these spectra can be described by the JP model (see Sect.~\ref{sec:JP}). Assuming a constant injection index throughout the lifetime of the source, the JP models can be jointly fit to these spectra, resulting in a precise measurement of the injection index optimised across the lobes. The integrated spectrum can now be fitted using the CI model for a known injection index, reducing any potential systematic errors in estimates for the break frequency and remnant ratio; the latter mapping directly to the duration of the remnant phase for any given source age. This is described in detail below. 

% Provided a sufficient spectral coverage, the CI model can be applied to the integrated spectra of radio AGNs in order to derive the break frequency (related to the spectral age), and crucially the fractional time spent within a remnant phase. Naturally this assumes the injection index is known, and while in principle it possible to simultaneously constrain all three parameters from the spectrum, this would require high-precision observations over a large enough frequency range so as to minimize the uncertainty and degeneracy in the modelled parameters (e.g. see \citealt{2018MNRAS.476.2522T}). This issue can be overcome by instead considering individual regions across the radio lobes; assuming a constant injection index throughout the lifetime of the source, the spectra arising from these regions will show various break frequencies (e.g. due to ageing of the plasma and structure in the magnetic field), however should be described by a common injection index. The injection index can therefore be modelled by jointly fitting these spectra using the JP models, which provides a way to estimate the value of $s$ that offers an optimal joint fit (e.g. see \citealt{2017A&A...600A..65S}). This method is explored below using observations of \remnant{}.

\begin{figure}
\centering
\includegraphics[width=\linewidth]{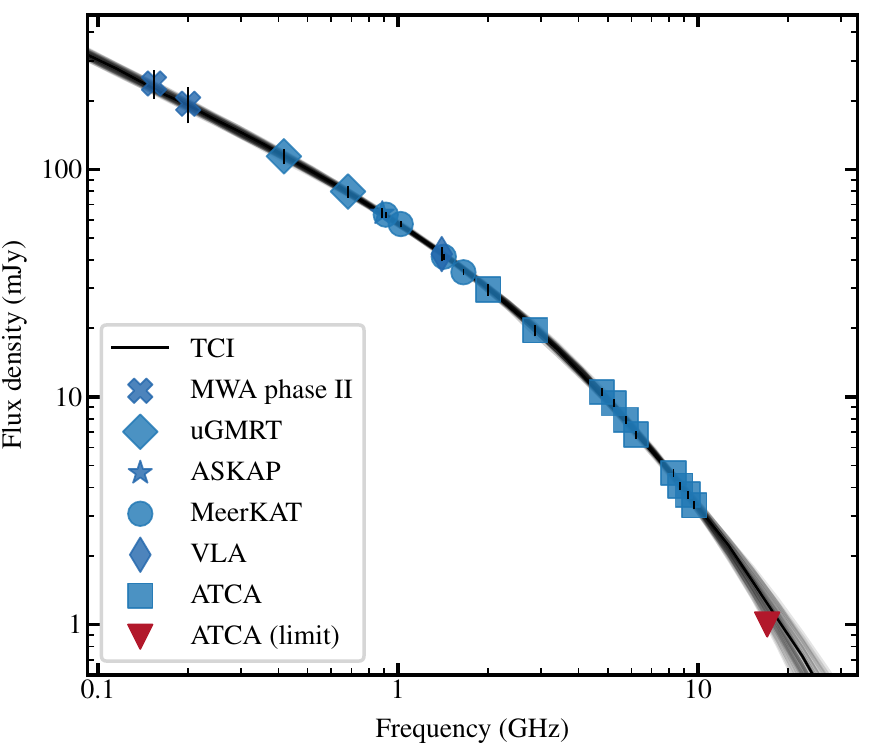}
\caption{The spectral emissivity of \remnant{}. Blue markers are used to display measurements of the integrated flux density. Error-bars represent 1$\sigma$ measurement uncertainties. Overlaid is the fit of the TCI model, assuming $B_0=0.1\,$nT and $s=2.25 \pm 0.03$ (Sect.~\ref{sec:injection_index}). The optimal fit of the TCI model (Sect.~\ref{sec:remnant_fraction}) is displayed using a solid black line. Also shown in light-gray are all plausible spectral fits within a 99.7\% ($3\sigma$) confidence interval. The spectral break frequency and remnant ratio are fitted as $\nu_\mathrm{b}~=~1.26_{-0.05}^{+0.06}$\,GHz, and $R_\mathrm{rem}=0.23 \pm 0.02$, respectively.}
\label{fig:integrated}
\end{figure}
\subsubsection{Modelling the injection index}
\label{sec:injection_index}

To model the injection index, we use the 8$''$ beam-matched radio maps described in Sect.~\ref{sec:data}. We use \textsc{miriad} to create a 2D cutout of \remnant{} from each image, and \texttt{regrid} the cutouts to have a common pixel scale ($d\theta=1''$) and projection (\texttt{SIN}). To efficiently sample the resolved radio spectrum, we extract the radio spectra arising from individual circular apertures arranged in a hexagonal configuration across the lobes. The full width half maximum (FWHM) of each aperture is set to $\theta_{\mathrm{FWHM}} = 8''$ to reflect the size of the restoring beam. The integrated flux density within each aperture is then calculated following $S_{\mathrm{int}} = \sum^n_{i=1}S_{p,i} \times d\theta^2/\Omega$, where $n$ is the number of pixels contained within each aperture, $S_{p,i}$ is the value of the $i^{\mathrm{th}}$ pixel in Jy\,beam$^{-1}$, $d\theta$ is the pixel angular scale, and $\Omega$ is the beam solid angle defined as $\theta_{\mathrm{FWHM}}^2\times\pi/(4\ln2)$. Given the spectra are extracted from exactly one resolution element, the measurement uncertainty is evaluated as the quadrature sum of the RMS noise in the observations and their absolute calibration uncertainty. 

Following this method, the radio spectra are extracted from all regions across the lobes with 9\,GHz radio emission above the 5$\sigma$ noise level. This restriction is done so as to maximise the number of observing frequencies contributing to the radio spectrum. The spectra arising from each aperture are independently fit using the JP (and TJP) models; importantly both the break frequency and the injection index are free parameters in their corresponding aperture spectral fit. In light of the results in Sect.~\ref{sec:field_independence}, we use Equation~2 of \citet{1980ARA&A..18..165M} to calculate a crude approximation of the lobe equipartition magnetic field strength as $B_{\mathrm{eq}} \approx 0.35\,$nT. Considering also that radio galaxy lobes are typically sub-equipartition \citep[e.g. see Fig.~11 of;][]{2018MNRAS.474.3361T}, we fit the TJP model assuming $B_0=0.1$\,nT, $0.35$\,nT and 1\,nT.
% Considering typical sub-equipartition values We use the results of \citet{2018MNRAS.474.3361T}, who fit the equipartition factors for a sample of 3C radio sources, and take their reported median value of $B/B_{\mathrm{eq}}=0.38$. In this way, we fit the TJP model assuming $B_0=0.132$\,nT.
% \begin{equation}
%     B_{\mathrm{eq}} = \Bigg[\frac{1+k}{(1+z)^{-\alpha-3}}\frac{S}{\nu^{-\alpha}\theta_x\theta_yl} \frac{\nu_{\mathrm{max}}^{0.5-\alpha}-\nu_{\mathrm{min}}^{0.5-\alpha}}{0.5-\alpha}\Bigg]^{\frac{2}{7}} [\mu G],
% \end{equation}
% where $k=1$
% As a precaution, we fit the TJP model assuming $B_0=0.1$\,nT, 1\,nT and 10\,nT, though expect this to have no influence on the derived values in light of the result discussed in Sect~\ref{sec:field_independence}. 
Spectral models are evaluated by uniformly sampling the injection index within a range corresponding to [2.01, 2.99] and an initial grid spacing of $\Delta s~=~0.05$. For each aperture spectral fit, a corresponding PDF is found by marginalising across the fitted injection index. Each PDF is then multiplied together to determine which value for the injection index provides an optimal fit across all aperture spectra; this assumes no correlation between the spectra arising from adjacent apertures, and also assumes a constant injection index throughout the lifetime of the source. Following this method, we optimise an injection index by considering the spectra arising from both lobes, which we present in Fig.~\ref{fig:injection}. 
% , which shows the result of jointly fitting these spectra in the south-west lobe using the TJP model and assuming a mean magnetic field strength of $B_0=0.1$\,nT. 
% Our results, optimised across both lobes, are summarised as follows: 
Importantly, we found that the optimal injection index modelled using the standard JP model is found to be $s~=~2.10\,\pm\,0.05$ (or; $\alpha_{\mathrm{inj}}=0.550\,\pm\,0.025$). For the Tribble model, we found an optimal value of $s~=~2.25\,\pm\,0.03$ (or; $\alpha_{\mathrm{inj}}=0.625\,\pm\,0.015$). 
% Marginal differences in the peak-probable injection index, fitted in each aperture, are likely due to the finite sampling 
% ; this is true for all values of $B_0$. 
% We find that the apertures closer to the projected radio center, which presumably contain plasma of an older spectral age, have less-certain confidence intervals centered about steeper values of the injection index. This is simply due to the finite sampling of a spectrum that has steepened appreciably even at low frequencies. Importantly, we find that these aged spectra are still consistent with the optimised injection index.
% Of the two spectral ageing models, we find that the TJP model offers a better model fit to the radio spectra. Nevertheless, heading into Sect.~\ref{sec:remnant_fraction} we consider both values for the injection index as plausible. 

\subsubsection{Modelling the break frequency \& remnant ratio}
\label{sec:remnant_fraction}
We compile an integrated radio spectrum for \remnant{} using the two decades of radio-frequency data described in Sect.~\ref{sec:data}. The raw spectrum is presented in Fig.~\ref{fig:integrated}, and demonstrates a clear spectral curvature owing to the losses associated with an inactive jet; this is evidenced by an ultra-steep high-frequency spectral index of $\alpha=1.7$, computed between 4.5\,-\,9.5\,GHz. With the injection index estimated in Sect~\ref{sec:injection_index}, we seek to model the break frequency and remnant ratio of \remnant{} by parameterising the observed curvature using the CI spectral ageing models. 

Similar to the resolved modelling to determine the injection index, we model the integrated spectrum using both the standard and Tribble forms of the CI model. For consistency, we use the injection indices estimated using the standard and Tribble JP models as inputs to the standard and Tribble CI models, respectively. Similarly, TCI models are evaluated for the same values of $B_0$ used in Sect.~\ref{sec:injection_index}. To account for the small uncertainty in the injection index previously fitted by the JP models, we treat `$s$' as a free parameter in the CI models, but apply a tight Gaussian prior based on the fitted uncertainty. We find that fixing the injection index offers no appreciable difference to our results, unsurprising considering the uncertainties on the injection index are small. The CI model fitted to the integrated spectrum is shown in Fig.~\ref{fig:integrated}.

Results of the fitting are summarised as follows (see also Tab.~\ref{tab:specmodelling}). Modelling the integrated radio spectrum using the CI model gives a break frequency of $\nu_\mathrm{b}=1.78_{-0.15}^{+0.17}$\,GHz, and a remnant ratio of $R_\mathrm{rem}=0.23 \pm 0.02$. Using instead the TCI model, gives a marginally different break frequency of $\nu_\mathrm{b}=1.26_{-0.05}^{+0.06}$\,GHz, and remarkably, a remnant ratio of $R_\mathrm{rem}=0.23 \pm 0.02$, consistent with the value constrained from the standard CI model.
% Similar to our finding in Sect~\ref{sec:injection_index}, we find that the values parameterised by the TCI model are consistent regardless the choice of $B_0$. 

\begin{table}
\centering
\caption{The spectral modelling attributes of \remnant{}, described in Sect.~\ref{sec:remnant_modelling}. Using the $8''$ radio images, the radio spectra arising from small regions across the lobes are jointly fit by the JP models, in order to optimise the injection index. The broadband integrated radio spectrum is then fit by the CI models, propagating the fitted injection index and uncertainties through a Gaussian prior, to estimate the break frequency and remnant ratio. }
\label{tab:specmodelling}
\begin{tabular}{lccc}
\toprule
Model & Injection index & Break frequency & Remnant ratio \\
\toprule
JP  & $2.10\,\pm\,0.05$ & $-$ & $-$ \\
TJP & $2.25\,\pm\,0.03$ & $-$ & $-$ \\
CI  & $-$ & $1.78_{-0.15}^{+0.17}$\,GHz & $0.23 \pm 0.02$\\[0.3em]
TCI & $-$ & $1.26_{-0.05}^{+0.06}$\,GHz & $0.23 \pm 0.02$\\
\bottomrule
\end{tabular}
\end{table}

Finally, it is important to mention that the precise measurements of the break frequency and remnant ratio of \remnant{} are only made possible by the independently-fitted injection index. By inspecting the marginal distribution of the fitted break frequency and remnant ratio, we find that their PDFs are Gaussian-like, and importantly do not demonstrate any parameter degeneracy worth consideration. These results are in stark contrast to those found by blindly fitting each parameter of the CI models from the integrated radio lobe spectrum. Here, we find that a single precise measurement of the injection index can not be found; e.g. for the standard CI model, the fitted injection index can plausibly range anywhere between $s\sim2.1$\,-\,$2.3$. Consequently, this introduces major uncertainties onto the fitted break frequency and remnant ratio, which become largely degenerate with one another; e.g. for the standard CI model, the break frequency demonstrates frequent local minima/maxima between $\nu_\mathrm{b}\sim2$\,-\,$5$\,GHz, and the PDF of the fitted remnant ratio is heavily asymmetric about its peak probable value of $R_{\mathrm{rem}}\sim0.38$. It is interesting that these results are found even for an integrated radio spectrum as well-sampled as that of \remnant{}, and thus reinforce the approach described in this section. 

\section{DYNAMICAL MODELLING OF REMNANT LOBES}
\label{sec:dynamical_modelling}
% \begin{figure*}
%     \centering
%     \includegraphics[width=1.02\textwidth,trim={0 4 0 8},clip]{RAiSEmaps.pdf}
%     \caption{Intensity maps predicted by RAiSE for a redshift $z=0.1$ radio source with a jet-power of $Q=10^{38}$\,W, source age of $\tau=100$\,Myr, and a remnant ratio of $R_\mathrm{rem}=0.25$. The radio source is shown for 150\,MHz (left~panels) and 1.4\,GHz (right~panels), each of which is shown for a native resolution corresponding to $1.024\times10^8$ particles (upper~panels) and for a convolved 10$''$ image (lower~panels). The pixel scale in the convolved image is $1''$. Following the method outlined in Sect.~\ref{sec:lobe_backflow_fitting}, the observed surface brightness distribution is optimally fit by a skewed two-dimensional Gaussian; this is shown by a solid black contour, which represents the edge bounding 95\% of the Gaussian volume. Images of each fitted Gaussian are truncated at the core, resulting in the apparent abrupt cut-off at 150\,MHz. }
%     \label{fig:RAiSEmaps}
% \end{figure*}

In the following section, we present a new dynamical model-based method that measures the off-time in remnant lobes, and is capable of identifying candidate remnant radio galaxies from observations. In Sect.~\ref{sec:raise}, we present the dynamical model that underpins this work. In Sect.~\ref{sec:lobe_backflow}, we justify why the backflow of lobe plasma holds valuable constraints on the off-time, and necessarily devise a tool to parameterise the surface brightness distribution of radio lobes. In Sect.~\ref{sec:radio_source_attributes}, we then outline the constraints necessary to model the energetics of lobed radio-loud \agns{}. Finally in Sect.~\ref{sec:modelling_the_energetics}, we implement our new method to constrain the energetics of \remnant{}, which we then independently verify against the remnant ratio previously measured in Sect.~\ref{sec:spectral_modelling}.

\subsection{Radio AGN in semi-analytic environments}
\label{sec:raise}
Throughout this work, the \textit{Radio AGN in Semi-analytic Environments} \citep[RAiSE;][]{2015ApJ...806...59T,2018MNRAS.473.4179T,2020MNRAS.493.5181T} code is used to model the dynamics of Fanaroff~\&~Riley type-II \citep[FR-II;][]{1974MNRAS.167P..31F} radio lobes. 
In this work, we employ RAiSE both as a tool to model the radio-loud AGN energetics, and to simulate mock radio source populations. 
% Here, we introduce the relevant details of the RAiSE model. 

\subsubsection{The RAiSE model}
\label{sec:raise_model}
 The RAiSE code consolidates models describing the expansion of lobed FR-II/I sources together with an observation-based treatment of the environments within which the jets expand; RAiSE has been tested to show consistency with hydrodynamical simulations, is able to reproduce the observed relationship between radio luminosity, morphology and host galaxy properties, is consistent with observed spectral ages maps of radio lobes, and is able to reconcile the discrepancy between spectral and dynamical ages in powerful radio galaxies. In this way, the temporal evolution of a radio source can be modelled for, amongst other intrinsic properties, any combination of source age, jet kinetic power, and intergalactic environment (forward modelling hereafter).

Quantifying an AGN environment requires knowledge of the distribution of gas in the associated dark matter halo. Ideally this is done by constraining a gas density and temperature profile through X-ray observations, however such observations are largely unavailable across the majority of haloes, especially at higher redshifts. \citet{2015ApJ...806...59T} showed that estimates of the halo mass of the intergalactic environment can alternatively be inferred from the optically-observed stellar content based on outputs of semi-analytic galaxy evolution models \citep[e.g. SAGE;][]{2006MNRAS.365...11C}; however, the shape of the gas density profile remains unconstrained by this method. RAiSE therefore quantifies the intergalactic environment in one of two ways: 1) by directly specifying a gas-density profile; or 2) by providing a mass of the dark matter halo, distributing the total gas mass based on an observationally-based gas density profile reported by \citet{2006ApJ...640..691V}. In the latter example, this can be done based on either the mean profile or by randomly sampling a distribution of environment profiles from a galaxy formation model.

RAiSE implements a lobe emissivity model that considers radiative losses due to synchrotron radiation and the inverse-Compton up-scattering of CMB photons, as well as the adiabatic losses related to the expansion of the lobes. This emissivity calculation is applied to small packets of electrons that are shock-accelerated at different points in the evolutionary history of the lobe. As a result, RAiSE considers the temporal evolution in the magnetic field; i.e. from the time electrons are shock-accelerated to the time they release their energy as synchrotron radiation. The magnetic field strength is derived from the lobe dynamics by assuming a constant ratio between the energy densities in the magnetic field and particles; referred to as the equipartition factor.

Over the past decade, models describing these processes have been incrementally added to the RAiSE code. \citet{2015ApJ...806...59T} provided the original implementation of a radio source model within a semi-analytic environment, and correspondingly modified the synchrotron emissivity model for the lobes. \citet{2018MNRAS.473.4179T} used hydrodynamical simulations to spatially map the synchrotron emissivity across the lobes, allowing them to produce surface brightness maps of simulated radio sources; their model implemented tracer particles from hydrodynamical simulations to probe the average locations and ages of injected particles. \citet{2018MNRAS.476.2522T} extended the dynamical and synchrotron emissivity models to account for the remnant phase during which the jets are inactive. Of relevance to this work, the dynamics presented by \citet{2020MNRAS.493.5181T} offer a slight modification to include the effect of a shocked gas shell.

\begin{figure*}
    \centering
    \includegraphics[width=1.02\textwidth,trim={0 4 0 8},clip]{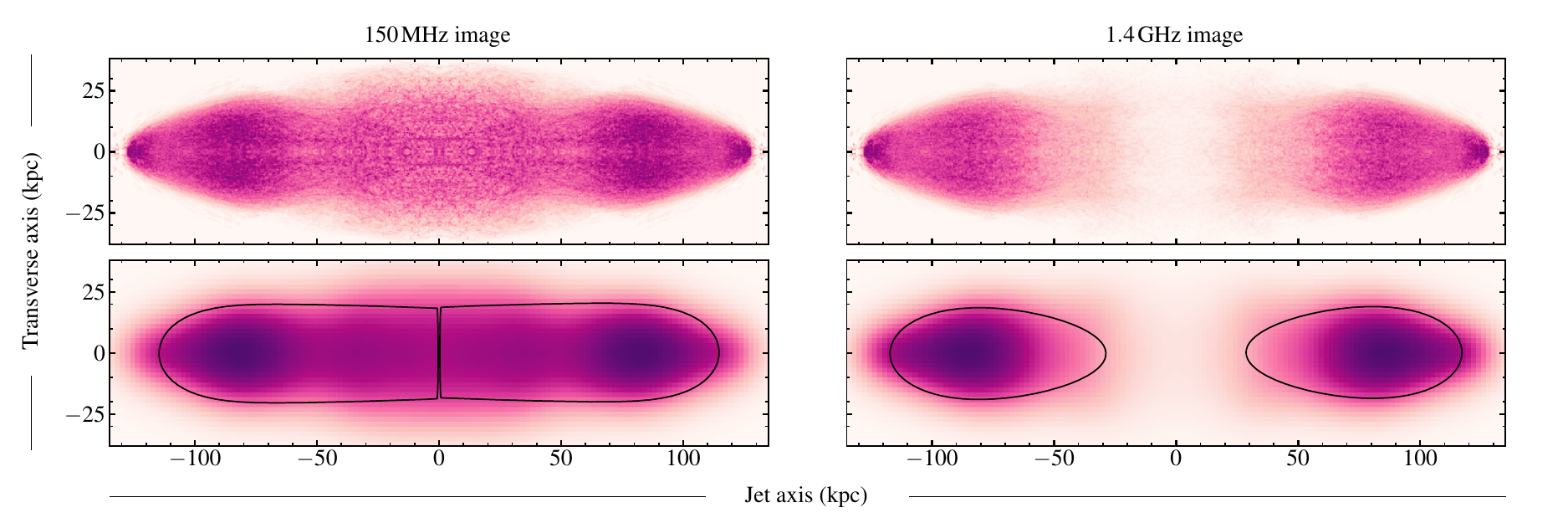}
    \caption{Intensity maps predicted by RAiSE for a redshift $z=0.1$ radio source with a jet-power of $Q=10^{38}$\,W, source age of $\tau=100$\,Myr, and a remnant ratio of $R_\mathrm{rem}=0.25$. The radio source is shown for 150\,MHz (left~panels) and 1.4\,GHz (right~panels), each of which is shown for a native resolution corresponding to $1.024\times10^8$ particles (upper~panels) and for a convolved 10$''$ image (lower~panels). The pixel scale in the convolved image is $1''$. Following the method outlined in Sect.~\ref{sec:lobe_backflow_fitting}, the observed surface brightness distribution is optimally fit by a skewed two-dimensional Gaussian; this is shown by a solid black contour, which represents the edge bounding 95\% of the Gaussian volume. Images of each fitted Gaussian are truncated at the core, resulting in the apparent abrupt cut-off at 150\,MHz. }
    \label{fig:RAiSEmaps}
\end{figure*}

The latest version of the RAiSE code\footnote{\url{https://github.com/rossjturner/RAiSEHD}}, as used in this work, is informed by the dynamics of \citet{2020MNRAS.493.5181T}, and implements the method of \citet{2018MNRAS.473.4179T} to produce intensity maps based on tracer particles from the higher-resolution hydrodynamical simulation of \citet{2022MNRAS.511.5225Y}. By using individual hydrodynamical simulation particles rather than population averages, this enables RAiSE to synthesise hydrodynamical-based, high-resolution maps of the radio lobes at any observing frequency (e.g. see Fig.~\ref{fig:RAiSEmaps}). 

\subsubsection{Constraining a RAiSE model}
\label{sec:constraining_a_raise_model}
The RAiSE model is characterised by a variety of model parameters that represent intrinsic parameters of a radio source. These parameters are constrained in one of three ways: 1) through a direct measurement (e.g. injection index); 2) a prior probability distribution based on measurements of well-studied objects (e.g. ambient density profile); or 3) by way of a parameter inversion based on simulated observable attributes (e.g. jet power and age from size and luminosity). 

% A key theme for this work is that the constraining power of the parameter inversion is limited by the input set of constraints.
RAiSE-based parameter inversions have been implemented by several previous authors to constrain the energetics of radio-loud \agns{}. \citet{2015ApJ...806...59T} demonstrated the ability to recover the jet powers and lifetimes of radio sources using only their observed physical size and monochromatic radio luminosity; in their work, halo masses for each source were inferred from the stellar masses of their hosts using SAGE, and intrinsic parameters such as the equipartition factor and injection index were fixed to typical values. \citet{2018MNRAS.474.3361T} were additionally able to recover the lobe equipartition factors by including the break frequency as an observational constraint. \citet{2020MNRAS.499.3660T} demonstrated that using only a reference flux density, angular size and width, break frequency and injection index, the cosmological redshift of radio sources can be constrained from the present epoch to the early-universe. 
% Despite needing an additional parameter to model the dynamics of remnant AGNs, \citet{2018MNRAS.476.2522T} demonstrated that the remnant-ratio can be independently derived from the integrated radio spectra of the remnant lobes, making their dynamical modelling no more complicated than active sources.

These results underline that the constraining power of the RAiSE-based parameter inversion is primarily limited by the input set of observational constraints. It is worth nothing that previous such applications of the RAiSE model have used attributes measured from observations with poor resolution. The capabilities of the current RAiSE model create a new opportunity to constrain intrinsic radio source parameters using additional attributes related to the observed surface brightness distribution of radio lobes. In the following sections, we exploit this opportunity to explore a new method for modelling the energetics of remnant \agns{}.

\subsection{The surface brightness distribution of radio lobes}
\label{sec:lobe_backflow}
In the absence of detailed spectral modelling, alternative constraints are needed to aid the modelling of the energetics of remnant \agns{}. Here, we propose that the backflow of plasma in radio galaxy lobes offers such constraints, and subsequently develop a tool to parameterise the observed surface brightness distribution of radio lobes. 
\subsubsection{Constraints embedded in the lobe backflow}
\label{sec:lobe_backflow_constraints}
The plasma contained within FR-II radio lobes is injected throughout the active phase at the hotspot, and is carried away from the hotspot by the backflow \citep[e.g.][]{2018MNRAS.473.4179T}. This leads to a gradient in age across the lobes, \textit{generally} in the form of an approximately linear increase in spectral age away from the hotspot. The observed shape of the emitting regions will vary depending on observing frequency; the low-frequency radiation produced by older electrons results in wider emitting regions along the jet-axis and transverse-axis, whereas at higher frequencies the bias towards observing younger electrons results in narrower emitting regions localised near the injection site. This is demonstrated in Fig.~\ref{fig:RAiSEmaps}. Importantly, the manner by which the observing frequency modifies the observed spatial distribution of synchrotron-emitting electrons, should depend on factors such as the plasma age distribution and lobe magnetic field strength. In the absence of spectral modelling, it is therefore reasonable to suggest that well-resolved observations of remnant radio lobes should offer the constraints necessary to help disentangle their active and remnant timescales.
\subsubsection{Skewed two-dimensional Gaussian lobes}
\label{sec:lobe_backflow_fitting}
We require attributes to describe the observed surface brightness distribution of radio lobes, which can be compared between observations and simulations. We note that the reproduction of fine details is neither important nor captured by the RAiSE simulations (e.g. knots in the jets, or slight bending of the jets). Implementing a skewed Gaussian in two spatial dimensions is appropriate for the following reasons: 1) a Gaussian is able to describe the fall-off in brightness as a function of position; 2) in two dimensions, a Gaussian is able to characterise a measure of length of the lobe along the jet-axis, as well as a width along the transverse-axis; 3) introducing a `skew' term along the major axis handles the asymmetry in brightness along the jet-axis (e.g. emission is typically brightest near the hotspot and falls off asymmetrically in each direction). In this way, the vastly different morphologies of, for example, a high-powered radio source (with a sharp fall-off in brightness away from the hotspot) and a low-powered aged remnant (with relatively smooth surface brightness profiles) can be characterised in a manner that is consistent.

To fit the skewed Gaussian to the observed surface brightness distribution of a radio lobe, the following free parameters are necessary: 
\begin{itemize}
    \item $S_0$, the amplitude of the Gaussian;
    \item $\mu_x$ \& $\mu_y$, the coordinates of the peak amplitude in each dimension;
    \item $\sigma_x$ \& $\sigma_y$, the standard deviation in each dimension;
    \item $\theta$, the major axis position angle;
    \item $\beta_x$, the skew term along the Gaussian major axis.
\end{itemize}
% We develop a small \textsc{python} package to implement this method. 
% the amplitude of the Gaussian, $\Phi_0$, the coordinates of the peak amplitude in each dimension, $\mu_x$ \& $\mu_y$, the standard deviation in each dimension, $\sigma_x$ \& $\sigma_y$, the major axis position angle, $\theta$, and the skew term, $\beta_x$. 
Since RAiSE simulates a jet in an known orientation and positions the SMBH at the center of the image, the angle of rotation ($\theta=0$) and $\mu_y$ are set by default (and therefore do not need solving). A simple least-squares algorithm is used to optimise the fit by minimising the residuals between the radio lobe surface brightness distribution and Gaussian model. For stability, regression is performed to the square-root of the input image; this ensures fitting to the general shape, and down-weights bias towards the brightest pixels. The resulting Gaussian fit can then be used to parameterise key attributes related to the observed surface brightness distribution of a radio lobe (see Sect.~\ref{sec:spatial_attributes} below). We make this code publicly available on GitHub\footnote{\url{https://github.com/benjaminquici/skewed-gaussian-lobes}}.
% An example of the fitting described in this section can be seen in Fig.~\ref{fig:RAiSEmaps}, which shows the Gaussian fitted to the surface brightness of a source simulated by the RAiSE model. 
% test

\subsection{Radio source attributes}
\label{sec:radio_source_attributes}
In this work, radio source attributes refer to the observable quantities that describe either the spatial, photometric, or spectral properties of radio galaxy lobes. By implementing these attributes in a RAiSE-based parameter inversion, they facilitate the comparison between observed radio sources and those simulated by the RAiSE model. However, images of real radio sources are `degraded' by the sensitivity and resolution of the observations; degrading simulated images with these same observing limitations therefore enables a more meaningful comparison. In the following section, we describe the methods for measuring these attributes from observed and simulated images. 

\begin{figure}
    \centering
    \includegraphics[width=1\linewidth]{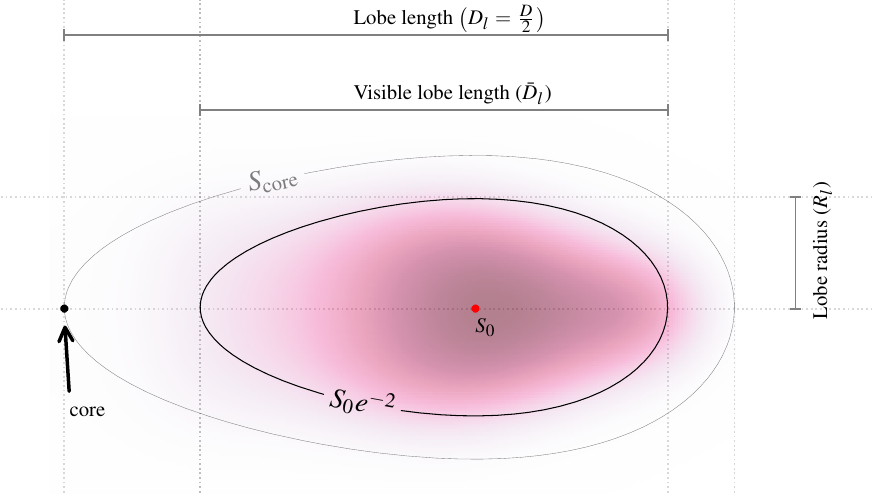}
    \caption{Schematic showing the key quantities necessary for computing the spatial attributes described in Sect.~\ref{sec:spatial_attributes}. The semi-transparent purple emission represents the right lobe of the 1.4\,GHz convolved image from Fig~\ref{fig:RAiSEmaps}. A skewed two-dimensional Gaussian is fit to the lobe surface brightness distribution following the method outlined in Sect.~\ref{sec:lobe_backflow_fitting}, where $S_0$ gives the amplitude of the fit. A solid black contour denotes the fitted Gaussian at a critical brightness level of $S_0e^{-2}$, and bounds 95\% of the Gaussian volume. }

    \label{fig:schematic}
\end{figure}

\subsubsection{Spatial attributes}
\label{sec:spatial_attributes}
Spatial attributes are measured based on the Gaussian fitting method described in Sect.~\ref{sec:lobe_backflow_fitting}. We do this by first considering the boundary of the 2D ellipse formed by slicing the fitted Gaussian at some critical brightness level (e.g. see Fig~\ref{fig:RAiSEmaps}). We define this critical level as $S_0e^{-2}$, above which 95\% of the Gaussian volume is contained. We stress that the arbitrary selection of this value does not matter; instead, what is important is that the same value is used when comparing both observed and simulated radio images, such that the resulting attributes are measured consistently. 
% 1) the physical size, $D$, defined as the largest linear size across the radio source; 2) the observed axis ratio, $\mathcal{A}_o$, defined as the ratio of the lobe length (parallel to the jet axis) over the lobe radius (orthogonal to the jet axis); and 3) a new attribute that we define as ``extent'', which considers how far back the emission sweeps towards the core. 
% It is important to highlight that the values measured for each spatial attribute must account for the telescope sensitivity/resolution. For example, the largest linear size of an observed radio source is typically measured out to the edge of the lobe
% For example the physical size, which for a prototypical FR-II is given by the separation between the jet termination points. While in principle this is a well-defined scale, this is not necessarily the case with real radio observations; the distance between the bright jet termination points depends on the telescope resolution as this will smear the true position of the terminal hotspots. Similar logic applies to the observed axis ratio and extent. The Gaussian-fitting method offers a convenient solution to this problem; so long as the simulated maps reflect the instrumental sensitivity and resolution, any attributes extracted are done so in a consistent manner. 
The following distances are measured by considering the boundary of the 2D ellipse (Fig.~\ref{fig:schematic}): the length of the lobe, $D_l$; the visible length of the lobe, $\bar{D_l}$; and the radius of the lobe, $R_l$. We consider the backflow in each lobe separately, and thus bound the visible lobe length as $0 < \bar{D_l} \leq D_l$. From here, we define four key spatial attributes, including the linear size, $D$, and observed axis ratio, $A$, as shown in Tab.~\ref{tab:spatial_attributes}. 

Importantly, to constrain the duration of the remnant phase we develop two new measures of ``extent'', which parameterise the observed backflow of lobe plasma. The ``extent~1'' attribute, $E$, considers what fraction of the total lobe length, parallel to the jet axis, is visible; e.g. how far back the lobe visibly sweeps towards the core. Lobes with emission localised near the hotspot will measure $E\rightarrow0$, which is expected if a majority of the lobe plasma energy is stored in the most freshly injected electrons. Conversely, $E\rightarrow1$ implies the observed radio emission sweeps all the way to (and beyond) the core, which is expected if a majority of the lobe plasma energy is stored in the older synchrotron-emitting electrons. The ``extent~2'' attribute, $\bar{E}$, considers how bright the lobes are at the core, relative to their peak brightness. We expect this attribute to be particularly useful at disentangling sources with backflow observed all the way to the core, where the measurement of the ``extent 1'' attribute will likely just reflect its upper limit of $E=1$. 
\begin{table}
\centering
\caption{Spatial attributes evaluated using the quantities shown in Fig.~\ref{fig:schematic}}
\label{tab:spatial_attributes}
\begin{tabular}{lccc}
\toprule
Spatial attribute & Symbol & Expression & Limit\\
\midrule
linear size & $D$ & $D_{l}^\mathrm{{lobe 1}} + D_{l}^\mathrm{{lobe 2}}$ & $-$\\
observed axis ratio  & $A$ & $D_l/R_l$ & $\geq1$\\
extent 1 & $E$ & $ D_l/\bar{D_l}$ & $0$\,-\,$1$\\
extent 2 & $\bar{E}$ & $S_\mathrm{core}/S_0$ & $0$\,-\,$1$\\

\bottomrule
\end{tabular}
\end{table}

% The linear size of the source can therefore be defined as sum of each lobe length: $D = D_{\mathrm{l,0}} + D_{\mathrm{l,1}}$. The axis ratio of a given lobe is defined as the ratio of the lobe length to the lobe radius: $A = D_l/R_l$. And similarly, the extent is defined as the ratio of the lobe length to the visible lobe length: $E = D_l/\bar{D_l}$. Characterizing extent in this way implies that $E\rightarrow1$ for sources with visible backflow near the core, and $E\rightarrow0$ for hotspot-dominated lobes. Since $\bar{D_l}$ can not exceed $D_l$, one caveat with this measure of extent is the loss of information for sources where the $\bar{D_l}$ is prematurely cut at the SMBH (e.g see 150\,MHz image of Fig~\ref{fig:RAiSEmaps}). To account for this, we define a second measure of extent, $\bar{E}$, that considers the ratio of the flux density fitted by the Gaussian at the AGN core, $S_\mathrm{core}$, to the peak flux density, $S_0$: $\bar{E} = S_\mathrm{core}/S_0$. This second measure of extent should be capable of distinguishing younger sources with varying levels of emission near the core, although will not be useful for older hotspot-dominated sources.

\subsubsection{Photometric attributes}
\label{sec:photometric_attributes}
A total radio luminosity, $L_\nu$, offers a powerful constraint on radio source energetics. For an observed radio source, the total radio luminosity can be derived from the total radio flux density by integrating over a region of the source bounded by a minimum brightness criterion; e.g. the region of the source above some signal to noise threshold. This means that, for a fixed sensitivity, the measured integrated flux density will depend on the surface brightness distribution of the lobes. The surface brightness maps produced by the RAiSE model allow us to include the effect of instrumental noise in the computation of the integrated flux density, thus accounting for this systematic bias. This is done by populating the simulated surface brightness maps with a Gaussian noise term, then the total observed radio luminosity is calculated considering only pixels brighter than $3\sigma$ (where $\sigma$ represents the level of Gaussian noise). This process of adding the noise is repeated over 16 iterations to sample a mean integrated flux density, and to characterise an uncertainty, $dL_{\nu}$, based on the standard deviation across each evaluation.

By measuring the integrated flux density over two observing frequencies, $\nu_0$~and~$\nu_1$, a two-point spectral index can be calculated through:
% \vspace{-1ex}
\begin{equation}
    \alpha~=~\log(L_{\nu,1}/L_{\nu,0})/\log(\nu_1/\nu_0),
\end{equation}
% \vspace{-1ex}
which has an uncertainty corresponding to:
\begin{equation}
    d\alpha~=~\big(1/\ln({\nu_0/\nu_1})\big)~\sqrt{(dL_{\nu,0}/L_{\nu,0})^2 + (dL_{\nu,1}/L_{\nu,1})^2}. 
\end{equation}
With at least three observing frequencies a spectral curvature, SPC, can be calculated through: $\mathrm{SPC} = \alpha_0 - \alpha_1$, which has an uncertainty corresponding to $d\mathrm{SPC}~=~\sqrt{(d\alpha_0)^2 + (d\alpha_1)^2}$.

\begin{table*}
\centering
\caption{The measured radio source attributes of \remnant{}, which are measured following the methods outlined in Sect.~\ref{sec:radio_source_attributes}. In Sect.~\ref{sec:bayesian_param_estimation} these attributes are compared with corresponding RAiSE model outputs, using the scale shown in column~4, to constrain the energetics via a parameter inversion. To explore the constraints embedded in the lobe backflow, attributes are compared in four different combinations (columns~5-8). Tick ($\checkmark$) and cross (\ding{55}) markers represent whether the attributes are included or excluded from the fitting, respectively.}
\label{tab:remnant_attributes}
\begin{tabular}{lccccccc}
\toprule
Radio source attribute & Symbol & Measurement & Scale & \textsc{full fit} & \textsc{spectral} & \textsc{spatial} & \textsc{in band} \\
\midrule
radio luminosity & $L_{1.4}$ & $(6.48\pm0.25)\times10^{24}$\,W\,Hz$^{-1}$ & $\log_{10}$ & $\checkmark$ & $\checkmark$ & $\checkmark$ & $\checkmark$\\[-0.15em]
$^{\text{(measured at 1.42\,\text{GHz})}}$ \\[-0.4em]
largest linear size & $D_{1.4}$ & 326\,kpc & $\log_{10}$ & $\checkmark$ & $\checkmark$ & $\checkmark$ & $\checkmark$\\[-0.15em]
$^{\text{(measured at 1.42\,\text{GHz})}}$ \\[-0.4em]
break frequency & $\nu_\mathrm{b}$ & $1.26^{+0.06}_{-0.05}$\,GHz & $\log_{10}$ & $\checkmark$ & $\checkmark$ & \ding{55}  & \ding{55} \\[+0.3em]
observed axis ratio & $A_{1.4}$, $A_{5.5}$ & $2.4\pm0.1$, $2.7\pm0.1$ & linear & $\checkmark$ & $\checkmark$ & $\checkmark$ & $\checkmark$\\[-0.15em]
$^{(1.42,~5.50\,\text{GHz})}$ \\[-0.4em]
extent 1 & $E_{1.4}$, $E_{5.5}$ & $0.9\pm0.05$, $0.8\pm0.05$ & linear & $\checkmark$ & \ding{55} & $\checkmark$ & \ding{55} \\[-0.15em]
$^{(1.42,~5.50\,\text{GHz})}$ \\[-0.4em]
extent 2 & $\bar{E}_{1.4}$, $\bar{E}_{5.5}$ & $0.76\pm0.05$, $0.65\pm0.05$ & linear & $\checkmark$ & \ding{55} & $\checkmark$ & \ding{55} \\[-0.15em]
$^{(1.42,~5.50\,\text{GHz})}$ \\[-0.4em]
spectral index & $\alpha_{150}^{1420}$ & $0.64\pm0.03$ & linear & \ding{55}  & \ding{55} & \ding{55} & $\checkmark$\\[-0.15em]
$^{(150-1420)\,\text{MHz}}$\\[-0.4em]
spectral curvature & $\mathrm{SPC}_{150}^{5500}$ & $0.53\pm0.04$ & linear & \ding{55} & \ding{55} & \ding{55}& $\checkmark$\\[-0.15em]
$^{(150-1420-5500)\,\text{MHz}}$\\[-0.4em]
\bottomrule
\end{tabular}
\end{table*}

\subsubsection{Spectral modelling attributes}
\label{sec:spectral_attributes}
The injection index and remnant ratio attributes, derived from observations by modelling the spectra of remnant radio galaxy lobes (Sect.~\ref{sec:spectral_modelling}), represent intrinsic parameters of the RAiSE model. On the other hand, a break frequency can be predicted by the RAiSE model, which offers valuable constraints on the lobe equipartition factor. While it is possible to measure a RAiSE-predicted break frequency by modelling a simulated radio spectrum, the computational cost of simulating a radio source at many different frequencies makes this method inefficient. For any given set of model parameters, the RAiSE code uses the lobe pressure and equipartition factor to compute the lobe magnetic field strength. In turn, the field strength is used through Eqn.~\ref{eqn:spectau} to calculate the corresponding RAiSE-predicted break frequency, where the redshift and source age are those used as RAiSE model inputs. By modelling the spectra of RAiSE-simulated radio sources, \citet{2018MNRAS.474.3361T} showed that the break frequency computed from the magnetic field strength is consistent with that derived by modelling their synthetic radio spectra (e.g. see their Fig.~3). 

\subsection{Modelling the energetics of \remnant{}}
\label{sec:modelling_the_energetics}
We are now equipped with the tools needed to model the energetics of \remnant{}. To begin with, we model the energetics closely following the dynamical model based method presented by \citet{2018MNRAS.476.2522T}. Here, the remnant ratio is already known (e.g. from Sect.~\ref{sec:remnant_fraction}), whilst the jet power, source age, and equipartition factor are principally constrained from a radio luminosity, physical size, and break frequency. We compare this established technique to our proposed method using the spatial attributes of the lobes (namely extent) to additionally constrain the remnant ratio. Finally, we provide our most precise constraints on the energetics by combining both the spectral and spatial attributes to best understand the physics of \remnant{}.

\subsubsection{Observed properties of \remnant{}}
\label{sec:observational_constraints}

The radio galaxy \remnant{} was classified as a remnant by \citet{2021PASA...38....8Q} based on the absence of radio emission from the core, and further supported by the observed steepening in the integrated radio spectrum. In that work, the radio source was unambiguously associated with a $z=0.2133$ host galaxy, for which a halo mass of $M_\mathrm{H}=10^{13.5}$\,M$_\odot$ was inferred based on the stellar mass of the host.

The radio source attributes of \remnant{} are measured following the methods outlined in Sect.~\ref{sec:radio_source_attributes}, and are summarised in Tab.~\ref{tab:remnant_attributes}. All spatial attributes are measured from the $8''$ images presented in Sect.~\ref{sec:data}, and we use the injection index, break frequency, and remnant ratio fitted in Sect.~\ref{sec:remnant_modelling} by the Tribble spectral ageing models (e.g. row~2~\&~4 of Tab.~\ref{tab:specmodelling}). In the following section, we implement these attributes to constrain the energetics of \remnant{} via a parameter inversion.

\begin{table}
\centering
\caption{The intrinsic parameters of \remnant{}, constrained via the four separate methods outlined in Tab.~\ref{tab:remnant_attributes}. \\
$^\dagger$ parameterised by the CI model.}
\label{tab:bayesian_parameters}
\begin{tabular}{ccccc}
\toprule
Param. & \textsc{full\,fit} & \textsc{spectral} & \textsc{spatial} & \textsc{in\,band} \\
\midrule
$Q$ & \multirow{2}{*}{$3.98^{+0.36}_{-0.40}$} & \multirow{2}{*}{$3.98^{+0.36}_{-0.40}$} & \multirow{2}{*}{$6.32^{+2.11}_{-1.63}$} & \multirow{2}{*}{$1.0$\,-\,$7.9$}\\[+0.15em]
$^{(\times10^{38}\text{ W})}$ \\[-0.55em]
$\tau$ & \multirow{2}{*}{$56.2^{+2.62}_{-1.28}$} & \multirow{2}{*}{$56.2^{+2.62}_{-1.28}$} & \multirow{2}{*}{$50.1^{+4.2}_{-3.8}$} & \multirow{2}{*}{$31$\,-\,$158$} \\[+0.00em]
$^{(\text{Myr})}$ \\[-0.30em]
$B$ & \multirow{2}{*}{$1.98^{+0.16}_{-0.15}$} & \multirow{2}{*}{$1.98^{+0.16}_{-0.15}$} & \multirow{2}{*}{$1.41^{+0.30}_{-0.25}$} & \multirow{2}{*}{$2$\,-\,$14$}\\[-0.05em]
$^{(\times10^{-10}\text{ T})}$\\[-0.35em]
$R_\mathrm{rem}$ & $^\dagger0.23^{+0.02}_{-0.02}$ & $^\dagger0.23^{+0.02}_{-0.02}$ & $0.26^{+0.04}_{-0.03}$ & $0.1$\,-\,$0.3$\\
\bottomrule
\end{tabular}
\end{table}

\subsubsection{RAiSE-based parameter inversion of \remnant{}}
\label{sec:bayesian_param_estimation}
RAiSE can model the temporal evolution of radio lobes for any set of intrinsic radio source parameters, and subsequently generate mock images to make predictions for their respective radio source attributes. The intrinsic parameters of any radio source can therefore be estimated via a RAiSE-based parameter inversion; observed radio source attributes are compared to a corresponding set of mock attributes generated over a multi-dimensional set of intrinsic parameters \citep[e.g as is done by;][]{2015ApJ...806...59T,2018MNRAS.474.3361T,2018MNRAS.476.2522T,2020MNRAS.496.1706S}.

For a fixed redshift of $z=0.2133$ and halo mass of $M_\mathrm{H}=10^{13.5}$\,M$_\odot$, a grid of RAiSE models are created from the two-sided jet kinetic power,~$Q$; the dynamical source age,~$\tau$; the equipartition factor,~$q$ \citep[which parameterises the ratio of energy densities in the magnetic field and particles;][]{2018MNRAS.473.4179T}; the initial supersonic phase value of the axis ratio,~$A_i$; the remnant ratio, $R_\mathrm{rem}$; and the injection index, $s$. We use the results of \citet{2018MNRAS.474.3361T}, who fit the intrinsic properties of 3C radio sources, to ensure our model grid covers plausible parameter values. In this work, RAiSE models are evolved between 10\,Myr to 1\,Gyr, recording outputs at every $\Delta(\log\tau)=0.05$\,dex. We uniformly sample $\log Q$ between $[37,40]$ with a precision of $0.05$\,dex; $\log q$ between $[-2.7,-0.7]$ with a precision of $0.05$\,dex; $A_i$ between $[2,8]$ with a precision of $0.25$; $R_\mathrm{rem}$ between $[0,5]$ with a precision of $0.01$; and $s$ between $[2.01,2.61]$ with a precision of $0.1$.

% Details of the parameter inversion are summarised Tab.~\ref{tab:bayesian_parameters}; the radio source attributes of \remnant{} are compared with their corresponding RAiSE model outputs
% Simulated images are simulated at 150\,MHz, 417\,MHz, 1.42\,GHz and 5.5\,GHz, and are degraded by the same set of observing conditions (sensitivity and resolution) as those 

To explore the constraining power of the extent attributes, we perform the parameter inversion based on four separate combinations of the radio source attributes, summarised in cols.~$5$\,-\,$8$ of Tab.~\ref{tab:remnant_attributes}. The \textsc{spatial} method tests a scenario where detailed spectral modelling (e.g. Sect.~\ref{sec:remnant_modelling}) can not be performed; here, each intrinsic radio source parameter is explicitly fit via the RAiSE-based parameter inversion, and the break frequency is omitted as a constraint. To independently verify results of this method, the fitted energetics are compared with that fitted by the \textsc{full~fit} and \textsc{spectral} methods; here, the injection index and, importantly the remnant ratio, are fixed to known values, and the strong constraint from the break frequency is included. The \textsc{in~band} method considers a limiting case where no constraints on the duration of the remnant phase are available. 

To conduct the parameter inversion (for any of the above methods), simulated images are used to measure the same attributes as those measured for \remnant{}, ensuring that the images are degraded by the same set of observing constraints. Attributes are compared with model outputs through Eqn.~\ref{eqn:likelihood}, and we follow the same process described in Sect.~\ref{sec:synchrofit} to select the best-fit model, and to estimate the uncertainty in each parameter.

The values fitted for the intrinsic parameters using each of these methods are presented in Tab.~\ref{tab:bayesian_parameters}. Consistent with the results of \citet{2018MNRAS.476.2522T}, we find that the jet-power/age/equipartition-factor triplet, fitted by the \textsc{spectral} method, is tightly constrained by the observed radio-luminosity/linear-size/break-frequency. We inspect the marginal distributions of each fitted parameter to find approximately Gaussian-like PDFs centered about the peak probable value; i.e. the fitting appears stable and not degenerate. By including constraints from the extents, we find that the energetics constrained by the \textsc{full~fit} method do not change. This suggests that the break frequency is the dominant observable in this fitting process. 

Using the source age constrained by \textsc{full~fit} method, together with the remnant ratio fit previously by the TCI model, the duration of the active and remnant phases in \remnant{} are calculated as $t_{\mathrm{on}}=43.3\pm4.3$\,Myr and $t_{\mathrm{rem}}=12.9\pm1.2$\,Myr, respectively. Importantly, through this method we demonstrate the ability to accurately measure the duration of the active phase, and to place a weak upper bound on the AGN jet duty cycle ($\delta\lesssim 1 - R_{\mathrm{rem}} = 0.77$). 
% We also note that for such a short duration remnant phase, the distribution of injected electrons is
% We note that for such a short duration remnant phase, the morpholog

% We note that for such a short duration remnant phase, and for a source with a linear size $>300$\,kpc, the buoyant rise of the remnant bubbles (not modelled by RAiSE) is insignificant, hence the model is perfectly valid.

Remarkably, we find that the energetics modelled by the \textsc{spatial} method are consistent with those derived by the previous methods. Uncertainties in each parameter are comparatively greater, as expected considering the constraints are relaxed and an additional free parameter is solved for, however inspecting the marginal distributions reveals similar Gaussian-like PDFs. What is particularly of interest is that the dynamically-modelled remnant ratio, $R_{\mathrm{rem}}=0.26\pm0.02$, is consistent with that parameterised by the TCI model within a level of $1\sigma$. Together with a source age of $\tau=50.1^{+4.2}_{-3.8}$\,Myr, the \textsc{spatial} method therefore estimates the duration of the active and remnant phases as $t_{\mathrm{on}}=37.2\pm4.3$\,Myr and $t_{\mathrm{rem}}=13.1\pm1.5$\,Myr, respectively. These timescales are consistent with those fitted by the \textsc{full~fit} method, suggesting that the surface brightness distribution of the lobe backflow can in fact be used to constrain the dynamics of the remnant phase.

Finally, we find that the stability of the model fitting breaks down severely with the \textsc{in~band} method. Inspecting the marginal distributions reveals a large degeneracy in the fitted parameter space, where multiple unique values of equal likelihood are found over a wide range of parameter values. We quote these ranges in Tab.~\ref{tab:bayesian_parameters}, by considering their minimum/maximum values. This tells us that the spectral index and curvature, alone, do not offer the constraining power needed to tightly constrain the energetics of \remnant{}.

% is able to tightly constrain a jet power of (); a total source age of (); and an equipartition factor of (), which corresponds to a lobe magnetic field strength of (). Our result shows that the jet-power/age/equipartition-factor triplet is tightly constrained by the radio-luminosity/linear-size/break-frequency, consistent with the results of \citet{2018MNRAS.476.2522T}. By comparison, we find that the \textsc{extent} algorithm fits a jet power of (); a total source age of (); an equipartition factor of (); and a remnant ratio of (). Although the parameters are fit with less precision, we find that the intrinsic parameters are consistent within the fitting uncertainties. Inspecting each marginalized parameter reveals approximately Gaussian distributions, suggesting the parameter are fit with stability. Remarkably, we find that the dynamically-modelled remnant ratio is consistent with that derived from the integrated radio spectrum. Finally, we find that the stability of the model fitting breaks down severely with the \textsc{no extent} algorithm. 
% By excluding the constraints from the extent, the equipartition 

\section{MOCK RADIO SOURCE POPULATIONS}
\begin{figure*}
    \centering
    \includegraphics[width=1\linewidth]{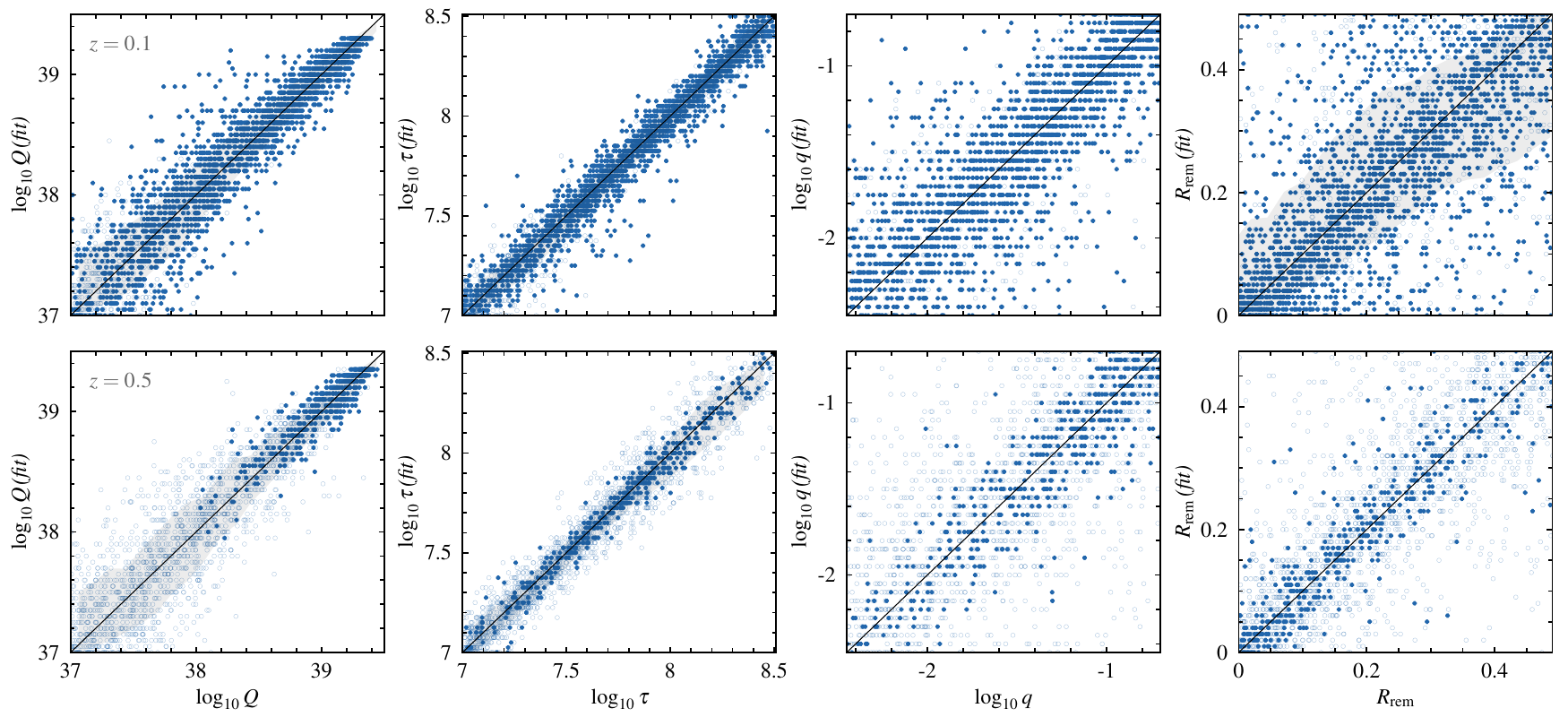}
    \caption{Results of the parameter inversion of the $z=0.1$ (upper row) and $z=0.5$ (lower row) mock catalogues are shown. Each plot demonstrates the fitted (vertical~axis) versus mock (horisontal axis) parameter for the jet power (first column), source age (second column), equipartition factor (third column), and remnant ratio (fourth~column). Each marker represents a unique mock radio source, where unfilled markers represent sources fainter than $S_{150}=100$\,mJy or smaller than $\theta=60''$. A solid black line is shown to represent a 1:1 relation between the fitted and mock quantities. }
    \label{fig:parameter_scatter}
\end{figure*}
\label{sec:mock_populations}

The ability to measure the off-time in remnant radio galaxies, using attributes measured exclusively below $\sim1.4$\,GHz, is particularly attractive considering the parameter space unlocked by wide-area sky surveys such as the LOFAR Two-metre Sky Survey \citep[LoTSS;][]{2019A&A...622A...1S,2022A&A...659A...1S}, the MeerKAT International GHz Tiered Extragalactic Exploration \citep[MIGHTEE;][]{2016mks..confE...6J} survey, the APERture Tile In Focus \citep[Apertif][]{2010iska.meetE..43O} survey, and EMU. However, the generalisability of this new method is currently unclear, as we do not know whether the constraining power of the spatial attributes, measured using telescope-degraded images, deteriorates for certain intrinsic source parameters; e.g. lower jet powers, older sources, or denser environments. In Sect.~\ref{sec:mock_catalogue}, we use our method to perform a parameter inversion of mock radio source populations simulated at 150\,MHz and 1.4\,GHz. In Sect.~\ref{sec:fitted_mock_parameters}, we compare the input and recovered parameters to examine the internal consistency of our method within the RAiSE model; e.g. does the method recover the input parameters. In Sect.~\ref{sec:remnant_selection} we then examine whether the remnant ratio fitted by this method can be used to confidently select between active and remnant radio lobes.

\subsection{Constructing mock radio source catalogues}
\label{sec:mock_catalogue}
For this analysis, mock radio source populations are simulated using the RAiSE model. We allow the following set of intrinsic parameters to vary over a wide range of values, typically associated with FR-IIs:

\begin{itemize}
    \item \emph{jet power:}\; $Q \in [10^{37},~10^{40}]\; \rm W$, to represent those typically associated with FR-II jets \citep{2015ApJ...806...59T}
    \item \emph{source age:}\; $\tau \in [10^7,~10^{8.5}]\; \rm yr$. Younger sources will likely fall below extended source criteria (see Fig.~\ref{fig:parameter_scatter}), and a genuine dearth of old ($>400$\,Myr) radio sources has been reported by \citet{2020MNRAS.496.1706S} in LOFAR data.
    \item \emph{remnant ratio:}\; $R_\mathrm{rem} \in [0,~0.5]$, consistent with the observational bias towards selecting recently switched-off remnants \citep{2018MNRAS.475.4557M,2018MNRAS.476.2522T,2020A&A...638A..34J,2020MNRAS.496.1706S,2021PASA...38....8Q}.
    \item \emph{equipartition factor:}\; $q \in [10^{-2.7},~10^{-0.7}]$, which is approximately Gaussian-distributed about $\log q \approx -1.7$ \citep[e.g. see;][]{2018MNRAS.474.3361T}.
    \item \emph{initial axis ratio:}\; A$_i \in [2.5,~4.5]$, reflecting a subset of typical values for the 3C sample \citep[e.g. see;][]{2015ApJ...806...59T}.
    \item \emph{injection index:}\; $s \in [2.01,~2.61]$, where the lower limit comes from predictions of first-order Fermi acceleration \citep{1962SvA.....6..317K,1970ranp.book.....P}, and the upper limit from $\alpha_{\mathrm{inj}}$ not being too steep. 
\end{itemize}

\begin{figure*}
    \centering
    \includegraphics[width=1\linewidth]{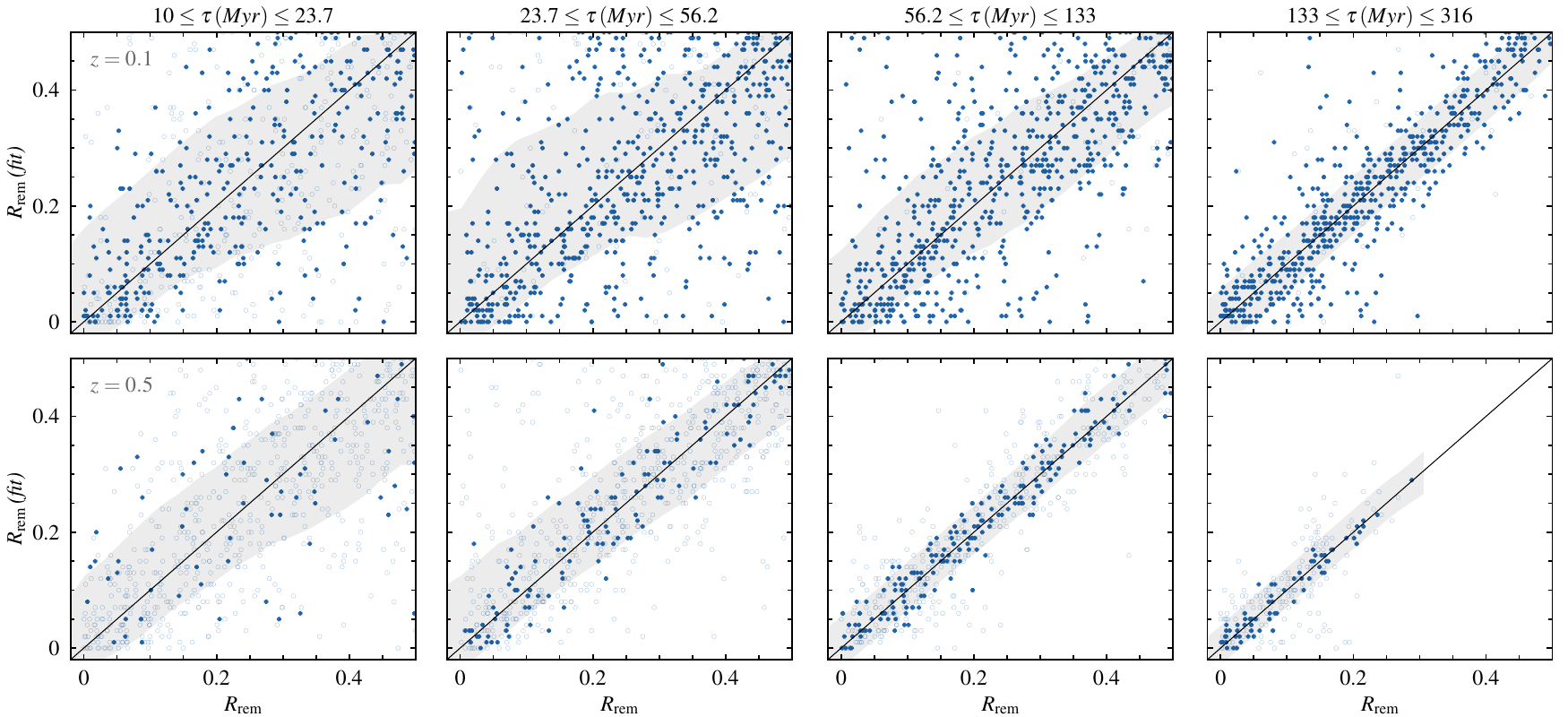}
    \caption{The fitted (vertical axes) versus input (horisontal axes) remnant ratio of the $z=0.1$ (upper panel) and $z=0.5$ (lower panel) mock catalogues. Each column represents a log-uniform source age bin. Gray regions represent a $1\sigma$ uncertainty in the scatter, corresponding to the unflagged sources (filled markers). A tightening in the scatter is seen as the source age increases, likely reflecting the greater magnitude of radiative losses discussed in Sect.~\ref{sec:fitted_mock_parameters}. The dearth of sources with higher values of $R_\mathrm{rem}$ in the $z=0.5$ sample arises due to the complete depletion of electrons capable of producing emission at these frequencies (e.g. see Fig.~4 of \citealt{2018MNRAS.476.2522T}).}
    \label{fig:rem_scatter}
\end{figure*}

Two separate mock catalogues are created at redshift $z=0.1,0.5$ (hereafter the $z=0.1$ and $z=0.5$ catalogues). For each catalogue, 5000 mock radio sources are forward modelled using pseudo-random values for the intrinsic parameters. We use a representative $M_\mathrm{H} = 10^{13}$\,M$_\odot$ halo mass (e.g. a typical environment of an FR-II, \citealt{2015ApJ...806...59T}), and simulate a random gas-density profile based on a prior distribution of environment profiles (as per discussion in Sect.~\ref{sec:raise_model}). 

A corresponding grid of RAiSE models is simulated with known values for each intrinsic parameter, so that a parameter inversion can be performed on each mock radio source catalogue. Our model grids are bound by the parameter space limits of the mock sample, however with uniform spacings as follows: $\Delta (\log Q)=0.05$\,dex; $\Delta (\log \tau)=0.025$\,dex; $\Delta R_{\mathrm{rem}}=0.01$; $\Delta (\log q)=0.05$\,dex; $\Delta A_i=0.25$; and $\Delta s=0.1$. Sources are simulated for the same halo mass, however here we take the mean of the environment profiles. 

Bayesian parameter estimation is conducted as follows. Images are synthesised at 150\,MHz and 1.4\,GHz, and degraded by a set of survey limitations shown in Tab.~\ref{tab:mock_observing_conditions}. This is done both for mock radio sources and those stored in the grid as model outputs. Telescope-degraded images are then used to measure the radio source attributes shown in Tab.~\ref{tab:mock_attributes}. Following the parameter inversion described in Sect.~\ref{sec:bayesian_param_estimation}, the attributes measured for mock radio sources are compared with those used as model outputs, in order to constrain their intrinsic parameters. Here, we simultaneously fit for the pseudo-randomly sampled parameters, and treat the redshift and halo mass as known. 

\begin{table}
\centering
\caption{The set of observing conditions used to sample mock radio source populations simulated in Sect.~\ref{sec:mock_catalogue}. }
\label{tab:mock_observing_conditions}
\begin{tabular}{lccc}
\toprule
\multirow{2}{*}{Survey type} & Freq. & Sensitivity & Resolution\\
 & (MHz) & ($\mu$Jy\,beam$^{-1}$) & ($''$) \\
\midrule
LOFAR-like &  150  & 70 & 6 \\[-0.2em]
$^{\emph{(wide-band)}}$ \\[-0.2em]
LOFAR-like & 130 & \multirow{2}{*}{100} & \multirow{2}{*}{6} \\[-0.2em]
$^{\emph{(sub-band)}}$& 170 &  &  \\[-0.2em]
% & 170 &  &  \\
MeerKAT-like & 1400 & 6  & 10 \\
\bottomrule
\end{tabular}
\end{table}

Sources are flagged if their angular size falls below $\theta=60''$, or if their 150\,MHz integrated flux densities fall below $S_{150}=100$\,mJy. This decision is motivated by the typical current observational constraints used to construct complete radio galaxy samples  \citep[e.g;][]{2017A&A...606A..98B,2017MNRAS.471..891G,2018MNRAS.475.4557M,2020A&A...638A..34J,2021PASA...38....8Q}. The results of this parameter inversion are explored in the following sections. 

Finally, we note that the Karl G. Jansky Very Large Array Sky Survey \citep[VLASS;][]{2020PASP..132c5001L} represents a high-frequency ($2-4$\,GHz), high-resolution ($2.''5$), all-sky (decl.$>-40^\circ$) radio survey, which is expected to achieve $\sigma=70\mu$Jy\,beam$^{-1}$ sensitivity. Despite marginally increasing the frequency leverage over 1.4\,GHz, we do not consider a VLASS-like observation due to the comparatively poorer surface brightness sensitivity to extended radio emission; without a handle over the fainter (older) lobe electrons, we expect the observed backflow to be much less constraining on the remnant off-time. 

% Our results should not be sensitive to the value prescribed for $M_H$. 

%  in this way, the uncertainty due to an unknown gas density profile is embedded in the fitted parameters of the mock radio source populations.

% here's how well we expect this to work given the noise in our parameter space
% not comparing if RAiSE predicts energetics correctlyt
% can RAiSE recover a jet power from a source, with whatever 
% RAiSE can predict Q and T, etc, 
% all you're needing to test is RAiSE is physically accurate
% what other real source do i expect this to work on

% For meaningful results, all images created by RAiSE are degraded by our observational constraints. 
% The following attributes are extracted from simulated surface brightness maps: a reference radio luminosity measured at 150\,MHz,~$L_{150}$; a low-frequency, in-band spectral index $\alpha_{130}^{170}$; a three-point spectral curvature measured as SPC\,=\,$\alpha_{130}^{170}~-~\alpha_{170}^{1400}$; a physical size measured at 150\,MHz,~$D_{150}$; and the remaining spatial attributes (axis ratio, and both measures of extent) are each measured at 150\,MHz and 1.4\,GHz. We perform a parameter inversion of each mock catalogue using attributes of the mock catalogue to compare with grid outputs; estimates of the best-fit parameters are made following the method outlined in Sect.~\ref{sec:bayesian_param_estimation}.

\subsection{Fitted energetics}
\label{sec:fitted_mock_parameters}
We expect that the parameter inversion may be less effective for some sets of parameters due to the non-linear application of radiative losses across our parameter space. 
In particular, the two extent attributes, and to a lesser degree the lobe width attribute, rely on the surface brightness distribution of ageing synchrotron-emitting lobe plasma. The ageing is quantified through the time-derivative of the Lorentz factor of a given packet of synchrotron-emitting electrons, $\gamma$, as follows \citep{1997MNRAS.292..723K}:
\begin{equation}
    \label{eqn:loss_rate}
    \frac{d\gamma}{dt} = -\frac{a_1\gamma}{3t} - \frac{4\sigma_\mathrm{T}\gamma^2\,(u_B+u_C)}{3m_ec},
\end{equation}
where the first and second terms encode adiabatic and radiative losses respectively. Here, the derivative is taken with respect to the age of the injected plasma, $t$, where $a_1$ gives the exponent with which the volume of the lobe grows adiabatically, $\sigma_\mathrm{T}$ is the cross section for Thompson scattering, and $u_B$ \& $u_C$ are the energy densities of the magnetic field and CMB respectively.

\begin{table}
\centering
\caption{The set of radio source attributes used to constrain the energetics of the mock radio source populations described in Sect.~\ref{sec:mock_catalogue}.}
\label{tab:mock_attributes}
\begin{tabular}{lcc}
\toprule
Attribute & Scale \\
\midrule
150\,MHz radio luminosity & $\log_{10}$ \\
largest linear size & $\log_{10}$ \\
observed axis ratio (150, 1400)\,MHz & linear \\
extent 1 (150, 1400)\,MHz & linear \\
extent 2 (150, 1400)\,MHz & linear \\
($150 - 1420$)\,MHz spectral index & linear \\
($150 - 1420$)\,MHz spectral curvature & linear \\
\bottomrule
\end{tabular}
\end{table}

Comparisons between their mock and fitted intrinsic parameters are presented in Fig.~\ref{fig:parameter_scatter} for the two mock catalogues described in Sect.~\ref{sec:mock_catalogue}. 
Firstly, although with varying degrees of confidence (as indicated by the scatter), we find that the fit of each parameter broadly closely matches the 1:1 relation with its mock equivalent. The jet power and source age are fit with greater confidence than the equipartition factor and remnant ratio, demonstrating the former two property's dependence on observable parameters; e.g. the jet power and source age are tightly constrained by the radio luminosity and linear size respectively, at least for active lobed sources \citep{2015ApJ...806...59T}. However, it is encouraging to find that the equipartition factor and remnant ratio are generally recovered as well. 

The jet power is less confidently predicted using the RAiSE-based parameter inversion for lower-powered sources, as shown in Fig.~\ref{fig:parameter_scatter}. This is unlikely a selection effect, in which the stochastic perturbations due to instrumental noise are more significant for fainter sources associated with lower-powered jets, since the same sizeable broadening is observed in the $z=0.1$ (brighter) catalogue. Instead, this is likely due to the strong correlation between jet power and the lobe magnetic field strength, and thus energy density $u_B$ (e.g. see Eqn.~A5 of \citealt{2018MNRAS.473.4179T}). The rate of energy loss from the synchrotron-emitting electrons is therefore expected to be much greater in higher jet power sources (Eqn.~\ref{eqn:loss_rate}), thus leading to a greater rate of change in the extent attributes as the remnant ages. Qualitatively this is consistent with the results of \citet{2018MNRAS.476.2522T}, who find a rapid dimming in higher-powered remnants in stark contrast to lower-powered sources whose flux density is maintained over a greater duration (see their Fig.~4). 

We also find that the energetics of the $z=0.5$ sample are fit with greater confidence (see also discussion in next paragraph). Again referring to Eqn.~\ref{eqn:loss_rate}, this result is unsurprising; the energy density of the CMB scales with redshift as $u_c \approxprop z^4$, meaning the magnitude of the energy losses sustained throughout the remnant phase will be greater. 

We investigate if the sizeable scatter in the remnant ratio is partially due to one of the other model parameters. In particular, we expect the radiative losses in the remnant phase will be related to the off-time, rather fractional time spent in the remnant phase (e.g. integrating Eqn.~\ref{eqn:loss_rate} for constant volume; i.e. $a_1 = 0$). As a result, we expect the synchrotron losses to reduce the extent attributes in approximate proportion to the time spent in the off-phase. The ability of the RAiSE-based parameter inversion to constrain the remnant ratio is therefore expected to correspondingly be much less effective in young sources (i.e. the off-time is smaller for a given remnant fraction).
To consider this potential sensitivity in our method, we group the mock catalogues into four source age bins uniformly-spaced in $\log \tau$. The fitted remnant ratio, in each source age bin, compared to the mock remnant ratio is presented in Fig.~\ref{fig:rem_scatter}. We find that the scatter in the predicted remnant ratio reduces considerably for the older source age bins, as expected from the above discussion.
\begin{figure}
    \centering
    \includegraphics[width=\linewidth]{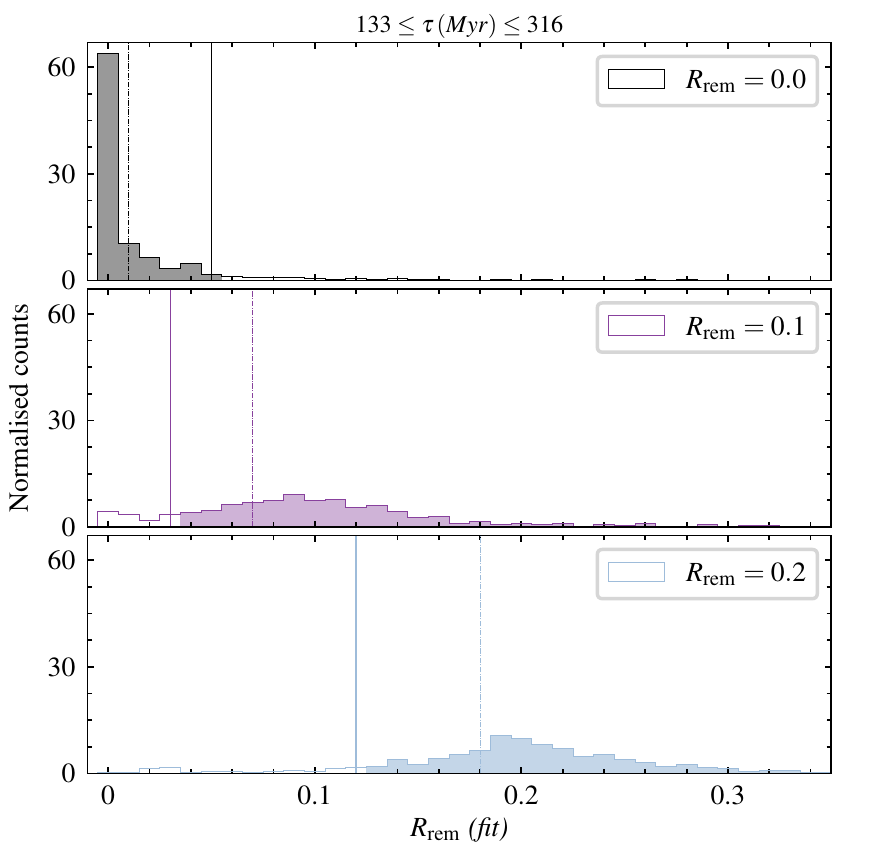}
    \caption{Normalised count representing the distribution of fitted remnant ratios of mock samples with $R_\mathrm{rem}=0$~(black), $R_\mathrm{rem}=0.1$~(purple), and $R_\mathrm{rem}=0.2$~(blue). We set the bin size in each distribution to $\Delta R_\mathrm{rem}=0.01$, to reflect spacings in the grid. Each sample is simulated at redshift $z=0.1$, and only the oldest source-age bin ($133~\leq~\tau~\leq~316$)\,Myr is shown. Solid and dotted vertical lines are shown to represent the 68$^{\mathrm{th}}$ and 95$^{\mathrm{th}}$ percentiles in each sample, respectively; these represent upper bounds for the active population, and lower bounds for the remnant samples. }
    \label{fig:lowTdistribution}
\end{figure}

Overall, the results presented here show promise for modelling the energetics of remnant AGNs, and notably, using spatial attributes of their lobes to constrain the remnant phase. The predictive power of this technique improves when the magnitude of the energy losses sustained during the remnant phase are greater; this is unsurprising, as the technique relies on the reduction in extent to constrain the ageing during this phase.
%To interpret the energetics fit by this method, one should first consider the age of the source to determine if the fitted remnant ratio will be estimated reliably.
The method makes reliable estimates of the remnant ratio associated with older sources, in contrast to young sources for which the fitted remnant ratio is less trustworthy.

\subsection{Identifying remnant radio galaxies}
\label{sec:remnant_selection}

\begin{table*}
\centering
\caption{The recovery rate of active and remnant radio sources simulated in Sect.~\ref{sec:remnant_selection}. Three mock radio source catalogues ($N=5000$) are created at $z=0.1$ and $z=0.5$, with remnant ratios fixed to $R_\mathrm{rem} = 0, 0.1, 0.2$, respectively. For each redshift (column~1) and source age bin (column~2), the $68^{\mathrm{th}}$ and $95^{\mathrm{th}}$ percentiles of the fitted remnant ratio distribution are reported (columns~$3-8$).}
\label{tab:contamination}
\renewcommand{\arraystretch}{0.99}
\begin{tabular}{cccrlrlrl}
\toprule
\multicolumn{2}{c}{Mock sample} & & \multicolumn{2}{c}{Active} & \multicolumn{2}{c}{Remnant $(R_\mathrm{rem}=0.1)$} & \multicolumn{2}{c}{Remnant $(R_\mathrm{rem}=0.2)$} \\

\textit{Redshift} & \textit{Source age (Myr)} & & (68$^{\mathrm{th}}$ \textit{perc.}) & (95$^{\mathrm{th}}$ \textit{perc.})& (68$^{\mathrm{th}}$ \textit{perc.}) & (95$^{\mathrm{th}}$ \textit{perc.}) & (68$^{\mathrm{th}}$ \textit{perc.}) & (95$^{\mathrm{th}}$ \textit{perc.})\\
\midrule
\multirow{4}{*}{$z=0.1$} & $10.0 \leq \tau < 23.7$ & & $\leq0.00$ & $<0.05$ & $>0.07$ & $>0.02$ & $>0.15$ & $>0.08$ \\
& $23.7 \leq \tau < 56.2$ & & $\leq0.00$ & $<0.12$ & $>0.07$ & $>0.01$ & $>0.15$ & $>0.06$ \\
& $56.2 \leq \tau < 133.0$ & & $<0.01$ & $<0.10$ & $>0.07$ & $>0.01$ & $>0.16$ & $>0.06$ \\
& $133.0 \leq \tau \leq 316.0$ & & $<0.01$ & $<0.05$ & $>0.07$ & $>0.03$ & $>0.18$ & $>0.12$ \\
\midrule
\multirow{4}{*}{$z=0.5$} & $10.0 \leq \tau < 23.7$ & & $\leq 0.00$ & $<0.05$ & $>0.06$ & $>0.01$ & $>0.14$ & $>0.03$ \\
& $23.7 \leq \tau < 56.0$ & & $\leq0.00$ & $<0.03$ & $>0.08$ & $>0.05$ & $>0.17$ & $>0.13$ \\
& $56.2 \leq \tau < 133.0$ & & $<0.01$ & $<0.03$ & $>0.09$ & $>0.06$ & $>0.19$ & $>0.17$ \\
& $133.0 \leq \tau \leq 316.0$ & & $\leq0.00$ & $<0.02$ & $>0.09$ & $>0.08$ & $>0.19$ & $>0.18$ \\
\bottomrule
\end{tabular}
\end{table*}

As we have demonstrated the ability of our method to estimate the remnant ratio with reasonable accuracy, this raises the question of whether we can use the same method to select remnant radio galaxies. Part of the difficulty here is ascertaining a definition for active and remnant sources based on the fitted remnant ratio parameter; logically, active sources would only correspond to $R_{\mathrm{rem}} = 0$ (i.e. zero off-time), however in practice it would be impossible to distinguish a very recently switched-off source from an active source. 

By creating mock radio source catalogues with known remnant ratios, we can use the distribution in the fitted remnant ratio to quantify the probability that: 1) an active radio source is fit with a non-zero remnant ratio; and 2) an ``unambiguous'' (i.e. $R_\mathrm{rem} \geq 0.1$) remnant radio source is fit with $R_\mathrm{rem}=0$. To do this, we follow the method outlined in Sect.~\ref{sec:mock_catalogue} to create three mock catalogues per redshift. Each catalogue contains 5000 sources, and the remnant ratios are fixed at $R_\mathrm{rem} = 0, 0.1, 0.2$, respectively. For each redshift and source age bin, we characterise the 68$^{\mathrm{th}}$ and 95$^{\mathrm{th}}$ percentiles of the fitted remnant ratio distribution; these serve as upper bounds for the $R_\mathrm{rem}=0$ catalogue, and lower bounds for the $R_\mathrm{rem}\geq 0.1$ catalogues. An example of this is demonstrated in Fig.~\ref{fig:lowTdistribution} for the highest source age bin at $z=0.1$, and we summarise our results in Tab.~\ref{tab:contamination} for all the redshift and source age bins. 

We find that the recovery rate of active radio galaxies is high; in $95\%$ of the cases, the fitted remnant ratio is $R_\mathrm{rem}\leq0.1$ (the actual value being 0\%), which is true across all redshift and source age bins (column~4 of Tab.~\ref{tab:contamination}). We conclude similar results for the $R_\mathrm{rem}=0.1$ remnant radio galaxies; in $95\%$ of the cases, the fitted remnant ratio is $R_\mathrm{rem}>0.01$ (column~6 of Tab.~\ref{tab:contamination}), meaning that active sources are not contaminated by $R_\mathrm{rem}=0.1$ remnants, at this level. For example, if examining a sub-catalogue of $z=0.1$ radio source with fitted ages of 133~to~316 Myrs, we can confidently (at $2\sigma$ level) say that those with remnant ratio $R_\mathrm{rem} < 0.03$ will not have significant contamination from ``unambiguous''-remnants and can therefore be classified as active sources. Conversely, those objects with fitted remnant ratios $R_\mathrm{rem} > 0.05$ will not have significant contamination from active sources and can be classified as remnants. We examine the same statistics for the $R_\mathrm{rem}=0.2$ sample, and reassuringly find no contamination of active sources from this population.

\section{CONCLUSIONS}
\label{sec:conclusion}
This paper investigates a new technique to constrain the energetics of remnant radio galaxies based on the backflow of their lobe plasma. We employ observations of the remnant radio galaxy \remnant{} together with mock radio source populations, in order to verify the method and examine its applications and limitations. The approach taken to explore this technique, and the insights gathered, are summarised below.

First, we use established spectral ageing techniques to derive a precise measurement of the off-time in \remnant{}; this is needed to independently validate the results of our new method. To do this, we collect new radio observations using the MeerKAT, ATCA and uGMRT telescopes (Sect.~\ref{sec:data}). The injection index is tightly constrained by jointly fitting the JP spectra arising from narrow regions across the lobes (Sect.~\ref{sec:injection_index}). We then make precise measurements of the break frequency and remnant ratio by fitting the TCI model to the integrated radio lobe spectrum (Sect.~\ref{sec:remnant_fraction}); here, the previously-fitted injection index is used as a model input. Our key results are summarised as follows: 
\begin{itemize}
    \item We develop a \textsc{python} implementation of synchrotron spectral ageing models, as well as algorithms to fit these models to observed radio data. We release this code, the \textsc{synchrofit} library, to the broader astrophysics community for its relevance to modelling the synchrotron spectra from active galaxies and supernovae remnants (Sect.~\ref{sec:synchrofit}).
    \item To place the tightest constraints on the break frequency and remnant ratio, the injection index should be independently fit and used as an input on the CI model; this can be done by considering individual JP spectra across the lobes \citep[cf.][]{2017A&A...600A..65S}.
    \item The shape of the Tribble spectral ageing models is (at most) weakly dependent on the average magnetic field of the underlying Maxwell-Boltzmann distribution (Sect.~\ref{sec:field_independence}). These models, which likely provide a more realistic treatment of a turbulent magnetic field, can therefore be fit to observed radio data without needing prior knowledge of the lobe magnetic field strength.
    \item The remnant ratio in \remnant{} is tightly constrained as $R_{\mathrm{rem}}=0.23\pm0.02$.
\end{itemize}

Our new method utilises the lobe surface brightness distribution to parameterise the backflow of plasma, and exploits the RAiSE dynamics and emissivity model. The high-resolution hydrodynamical simulation underpinning the RAiSE code enables the synthesis of mock surface-brightness maps, which are used to measure predicted radio source attributes. These are compared with observations so that the intrinsic source properties can be obtained via a parameter inversion. We apply this method to constrain the energetics of \remnant{}, and obtain the following results:
\begin{itemize}
    \item We propose two new attributes to quantify the spectral ageing of lobe plasma as it flows away from the hotspot towards the core. The ``extent 1'' attribute considers how far back the lobe visibly sweeps towards the core, and the ``extent 2'' attribute considers the surface brightness of the lobe at the core, relative to the peak surface brightness.
    \item We implement a RAiSE-based parameter inversion using these two new attributes, in addition to the lobe radio luminosity, size, and axis ratio, to constrain a remnant ratio of $R_\mathrm{rem} = 0.26\pm0.02$ for \remnant{}. Importantly, this is consistent at the $1\sigma$ level with that constrained by our spectral fitting, verifying our approach to constrain the dynamics of the remnant phase based on the surface brightness distribution of the lobe backflow.
    \item By comparing different sets of attributes, we find that the break frequency is the dominates the ``extent'' attributes fitting of the energetics; particularly the equipartition factor and remnant ratio. By excluding the break frequency and extent attributes from the fitting, we find that the spectral index and curvature, alone, do not offer enough constraining power to uniquely fit the source energetics.
    \item For \remnant{}, we accurately measure a two-sided jet kinetic power of $Q = 3.98^{+0.36}_{-0.40}\times10^{38}$\,W, a lobe magnetic field strength of $B = 0.20\pm0.02$\,nT, and importantly an active and remnant timescale of $t_{\mathrm{on}}=37.2\pm4.3$\,Myr and $t_{\mathrm{rem}}=13.1\pm1.5$\,Myr.
\end{itemize}

We then used synthetic radio lobes, simulated for a typical 150\,MHz LOFAR-like and 1.4\,GHz MeerKAT-like observation, to further examine the limitations of our proposed method with respect to intrinsic radio source parameters. Our key results are summarised as follows: 
% In spirit of the results above, we investigate how well the input parameters of mock radio sources, simulated by the RAiSE model, can be recovered over a wide range of intrinsic parameters. A population of 5000 mock sources is simulated at $z=0.1$ and $z=0.5$. Attributes are extracted from mock surface-brightness maps assuming a 150\,MHz LOFAR-like observation, and 1.4\,GHz MeerKAT-like observation. Using these attributes, the input parameters of the mock radio sources are estimated via a RAiSE-based parameter inversion. Our key results are summarised as follows: 
\begin{itemize}
    \item The jet power, source age, equipartition factor, and remnant ratio are consistently recovered, noting that the former two are constrained with high precision. This result builds on the work conducted by \citet{2018MNRAS.474.3361T} for active sources only, who similarly find that the jet power, source age, and equipartition factor are constrained from the source size, radio luminosity, and break frequency.
    \item Our results show that that the parameters are more accurately fit for radio sources with: higher redshifts, due to increased strength of IC losses; higher-powered jets, due to increase in magnetic field strength; and for a greater duration of the remnant phase. We therefore find that the constraining power of our method increases when the magnitude of the energy losses, sustained during the remnant phase, are greater.
    \item Remnant ratios fitted by this method can be trusted if the measured source age is large, and conversely, is less accurate for younger sources. To interpret the on/off-times fit by this method, one should first consider the age of the source to determine if the fitted remnant ratio is estimated reliably. 
    \item We find that our method is capable of selecting active and remnant radio galaxy candidates. At low redshift ($z\sim0.1$), we find that sufficiently aged sources ($\tau>50$\,Myr) fitted with $R_{\mathrm{rem}}\leq0.1$ are active radio galaxies within a confidence of $95\%$; this extends to sources as young as $\tau\sim10$\,Myr at higher redshifts ($z\gtrsim0.5$). Similarly, we find that sources fit with $R_{\mathrm{rem}}\geq0.1$ are true remnant radio galaxies within a confidence of $95\%$; these classifications are more robust at higher redshift. 
\end{itemize}

The outcomes of this work show promising applications to studying the energetic impact of radio-loud AGNs on their environments, in particular, through studying remnant radio galaxies. %The ability of this technique to tightly constrain, in particular, the jet kinetic power and the duration of the active phase, should be exploited across a larger sample of remnant radio-loud AGNs; the results of our work suggest that this can be carried using only a low and intermediate observing frequency, provided a sufficiently high ($\approx10''$) spatial resolution. 
However, although this method is clearly successful for remnants with $R_{\mathrm{rem}}\lesssim 0.23$, for much older remnants the model assumptions in the version of RAiSE used in this work may no longer capture all the necessary physics, for example buoyancy and Rayleigh-Taylor instability; capturing these physical processes will be an important step necessary for modelling aged remnants, e.g. Blob~1 with $R_{\mathrm{rem}}\sim 0.8$. Other complexities, such as dynamic or non cluster-centered environments, can lead to asymmetries in the radio lobe morphologies; e.g. in jet lengths, as well as lobe luminosities. Hydrodynamical simulations of the jet and lobe evolution in such environments already exist \citep[see;][]{2021MNRAS.508.5239Y}, which can incorporate the calculation of the resolved adiabatic and radiative loss processes following \citet{2022MNRAS.511.5225Y}. While the direct outputs of such hydrodynamical simulations can be implemented in the grid-based approach of this work, their synthesis is computationally expensive and can realistically only coarsely sample any given parameter space. The RAiSE framework can naturally consider asymmetries by allowing its governing differential equations for the source expansion to be applied on arbitrarily small solid angle elements; in this manner, future work can utilise modified versions of this code that include relevant physical processes to efficiently simulate millions of mock sources densely throughout the intrinsic parameter space.

\section*{ACKNOWLEDGEMENTS}
BQ acknowledges a Doctoral Scholarship and an Australian Government Research Training Programme scholarship administered through Curtin University of Western Australia. NHW is supported by an Australian Research Council Future Fellowship (project number FT190100231) funded by the Australian Government. We acknowledge the Pawsey Supercomputing Centre which is supported by the Western Australian and Australian Governments. The Australia Telescope Compact Array is part of the Australia Telescope National Facility which is funded by the Australian Government for operation as a National Facility managed by CSIRO. We thank the staff of the GMRT that made these observations possible. GMRT is run by the National Centre for Radio Astrophysics of the Tata Institute of Fundamental Research. 
% We acknowledge the work and support of the developers of the following following python packages: Astropy \citep{astropy:2013, astropy:2018} and Numpy \citep{vaderwalt_numpy_2011}. This work was compiled in the very useful free online \LaTeX{} editor Overleaf.
We additionally would like to thank Aleksandar Shulevski for his helpful suggestions that have improved our manuscript.

\section*{DATA AVAILABILITY}
The raw data products underlying Sect.~\ref{sec:data} are available on the following data archives: the Australia~Telescope~Online~Archive~(ATOA) at \url{https://atoa.atnf.csiro.au/} (Sect.~\ref{sec:atca}); the South African Radio Astronomy Observatory (SARAO)~Web~Archive at \url{https://archive.sarao.ac.za/} (Sect.~\ref{sec:data:meerkat}); and the GMRT Online Archive (GOA) at \url{https://naps.ncra.tifr.res.in/goa/data/search} (Sect.~\ref{sec:gmrt}). Processed data products underlying this article will be shared on reasonable request to the authors.
Access to the code-bases used throughout this work are disclosed within the paper. 
% throughout this This work also makes use of several code-bases, available publicly on GitHub: the \textsc{polygon\_flux} code used in Sect.~\ref{sec:data} is available on \url{https://github.com/nhurleywalker/polygon-flux}; the \textsc{synchrofit} code developed in Sect.~\ref{sec:synchrofit} is available on \url{https://github.com/synchrofit}; the RAiSE code described in Sect.~\ref{sec:raise_model} is available on \url{https://github.com/rossjturner/RAiSEHD}; the skewed Gaussian fitting code developed in Sect.~\ref{sec:lobe_backflow_fitting} is available on \url{https://github.com/benjaminquici/skewed-gaussian-lobes}. 
% This paper makes uses of several original data generated over the course of this study. The 
% Processed data products underlying this article will be shared on reasonable request to the authors.
% The data underlying this article are available in [repository name, e.g. the GenBank Nucleotide Database] at [URL], and can be accessed with [unique identifier, e.g. accession number, deposition number].
% The data underlying this article will be shared on reasonable request to the corresponding author.

\appendix
\setcounter{section}{1}
\label{sec:appendix}
\begin{table*}
\begin{center}
\caption{Summary of the integrated radio spectrum compiled for \remnant{}.}
\label{tab:integrated_flux}
\renewcommand{\arraystretch}{0.9}
\begin{tabular}{lcccc}
\toprule
\multirow{2}{*}{Telescope} & \multirow{2}{*}{Type} & Frequency & Bandwidth & Integrated flux density\\
&  & (MHz) & (MHz) & (mJy) \\
\toprule
\multirow{2}{*}{MWA Phase I} & \multirow{2}{*}{Survey (GLEAM)} & 119.04 & \multirow{2}{*}{30.72} & $238 \pm 35$ \\
& & 154.88 & & $194 \pm 35$ \\
\hline 
\multirow{2}{*}{uGMRT} & \multirow{2}{*}{Pointed observations} & 417 & \multirow{2}{*}{250} & $114.00 \pm 8.88$ \\
& & 682 & & $79.8 \pm 5.2$ \\
\hline
ASKAP & Survey (EMU early-science) &  887 & 288 & $64.1 \pm 3.2$ \\
\hline
\multirow{4}{*}{MeerKAT} & \multirow{4}{*}{Pointed observations} & 911 & 42 & $63.17 \pm 1.89$ \\
& & 1022 & 120 & $57.44 \pm 1.85$ \\
& & 1422 & 153 & $41.28 \pm 1.34$ \\
& & 1653 & 30 & $35.28 \pm 1.08$ \\

\hline
VLA & Survey (NVSS) & 1400 & 50 & $42.4 \pm 3.05$ \\
\hline
\multirow{10}{*}{ATCA} & \multirow{10}{*}{Pointed observations} &2000 & \multirow{10}{*}{500} & $29.58 \pm 1.77$ \\
& & 2868 & & $19.60 \pm 1.17$ \\
& & 4782 & & $10.50 \pm 0.42$ \\
& & 5243 & & $9.38 \pm 0.37$ \\
& & 5746 & & $7.93 \pm 0.32$ \\
& & 6216 & & $6.84 \pm 0.27$ \\
& & 8284 & & $4.13 \pm 0.19$ \\
& & 8726 & & $3.84 \pm 0.16$ \\
& & 9253 & & $3.41 \pm 0.15$ \\
& & 9716 & & $3.08 \pm 0.14$ \\
 
\bottomrule
\end{tabular}
\end{center}
\end{table*}
% \begin{table*}
% \begin{center}
% \caption{Summary of the integrated radio spectrum compiled for \remnant{}.}
% \label{tab:integrated_flux}
% \renewcommand{\arraystretch}{0.9}
% \begin{tabular}{lcccccccccccccccccccc}
% \toprule
% Telescope \\
% Frequency (MHz) & 119.04 & 154.88 & 417 & 682 & 887 & 911 & 1022 & 1422 & 1653 & 1400 & 2000 & 2868 & 4782 & 5243 & 5746 & 6216 & 8284 & 8726 & 9253 & 9716 \\
% Bandwidth (MHz)\\
% Integrated flux density \\
% \bottomrule
% \end{tabular}
% \end{center}
% \end{table*}
\bibliographystyle{pasa-mnras}
\bibliography{main.bib}
\end{document}